\documentclass[letter, 11pt]{article}
\usepackage{natbib}
\usepackage{latexsym}
\usepackage{graphics}
\usepackage{amsfonts}
\usepackage{enumerate}
\usepackage{graphicx}
\usepackage{epsf,boxedminipage,lscape}
\usepackage{epstopdf}
\usepackage{epsfig}
\usepackage{verbatim}
\usepackage{subfig}
\usepackage{caption}
\usepackage{footnote}
\usepackage{slashbox}
\usepackage{lscape}
\usepackage{hhline}
\usepackage{array}
\usepackage{graphicx}
\usepackage{setspace}
\usepackage{amssymb}
\usepackage{amsmath}
\usepackage{amsfonts}
\usepackage{times}
\usepackage{color}
\usepackage{amsthm}
\usepackage{indentfirst}
\usepackage{epsfig}
\usepackage{lscape,ctable,rotating}
\usepackage{enumitem}
\setlist{nolistsep}
\usepackage{rotating}
\usepackage[left=1in,top=1in,right=1in,foot=1in]{geometry}
\usepackage[ruled,vlined]{algorithm2e}

\newtheorem{definition}{Definition}[section]

\newtheorem{exam}[definition]{Example}

\newtheorem{rem}{Remark}[section]

\newcommand{\R}{\mathbb{R}}

\newcommand{\var}{\textrm{var}}

\newcommand{\levy}{{L\'evy}}

\newcommand{\V}{{\mathcal{V}}}

\newcommand{\disteq}{\stackrel{\mathrm{d}}{=}}

\newcommand{\dt}{{\varDelta t}}

\title{Sample path generation of the stochastic volatility CGMY process and its application to path-dependent option pricing}
\author{Young Shin Kim\footnote{College of Business, Stony Brook University, New York, USA (aaron.kim@stonybrook.edu) }}

\begin{document}
\doublespacing
\maketitle
 
\begin{abstract}
This paper proposes the sample path generation method for the stochastic volatility version of CGMY process. We present the Monte-Carlo method for European and American option pricing with the sample path generation and calibrate model parameters to the American style S\&P 100 index options market, using the least square regression method. Moreover, we discuss path-dependent options such as Asian and Barrier options.
\end{abstract}

\section{Introduction}

Since \cite{Heston:1993} applied the CIR model by \cite{CIR1985} to option pricing, the model has been the standard framework for option pricing because it allows for stochastic volatility and volatility smile effect that observed for the Black Scholes model (\cite{BlackScholes:1973}). However, the \levy~process models with time-varying volatility have been used in option pricing in descrete time model, since empirical studies based on the stochastic volatility model show that the Brownian Motion is often rejected  (See \cite{RachevMittnik:2000}, \cite{Kim_et_al:2010:JBF}, \cite{Kim_et_al:2011}). \cite{Carr_et_al:2003} defined the class of continuous time stochastic volatility model on \levy~processes (SVLP) using the time changed \levy~process. The SVLP has been successfully applied in European option pricing, but the absence of an efficient sample path generation method makes the SVLP model hard to be applied to path-dependent options such as American, Barrier or Asian option.

This paper proposes the sample path generation method for the stochastic volatility version of CGMY (CGMYSV) process, which is a subclass of the SVLP model. The method is constructed by an approximation of the series representation of CGMY process with the time-varying scale parameter. The series representations for the tempered stable and tempered infinitely divisible processes are discussed in \cite{Rosinski:2007} and \cite{Bianchi_et_al:2010}, and it is applied to Monte-Carlo simulation for CGMY market model with a GARCH volatility in \cite{Kim_et_al:2010:JBF}. Yet, the CGMYSV model is a continuous-time model different from the GARCH model with CGMY innovations. We develop an algorithm of the CGMYSV sample path generation, and it will be applied Monte-Carlo simulation (MCS). The algorithm will be used to price European and American option and to calibrate risk-neutral parameters to the S\&P 500 index option (European style) and S\&P 100 option (American style) data. We will use the least square regression method by \cite{Longstaff_Schwartz:2001} for American option pricing with MCS. We verify that the new sample path generation method performs well in American option pricing by that empirical study. We also apply the algorithm to Asian and Barrier option pricing with MCS.

The remainder of this paper is organized as follows. 
The CIR process and the CGMY process with the series representation are presented in Section 2. The sample path generation method based on the series representation is constructed in Section 3. In Section 4, we perform the CGMYSV model calibration for S\&P 500 index option and S\&P 100 index option. Also, the Asian and Barrier option prices are discussed. Finally, Section 6 concludes. 

\section{Preliminary}
In this section, we briefly discuss CIR model and CGMY process.
\subsection{\label{sec CIR Process} CIR model}
The CIR model is given by
\begin{equation}\label{eq CIR Process}
dv_t = \kappa(\eta-v_t) dt + \zeta\sqrt{v_t} dW_t \text{ and } v_0>0,
\end{equation}
for $\kappa, \eta, \zeta >0$ and Brownian motion $\{W_t\}_{t\ge 0}$. 
Let $\mathcal F^v_{t}$ be a smallest $\sigma$-algebra generated by process $\{v_{s}\}_{0\le s \le t}$ then $v_{t+\dt}|_{\mathcal F^v_{t}} \disteq \xi/(2c)$ where $c = \frac{2\kappa}{(1-e^{-\kappa \dt})\zeta^2}$ and the random variable $\xi$ is non-central $\chi^2$-distributed with degrees of freedom $4\kappa\eta/\zeta^2$ and noncentrality parameter $2cv_te^{-\kappa \dt}$.

Let $\V_t = \int_0^t v_s ds$ then the joint distribution of $\left(v_t,, \V_t\right)$ is characterized by the characteristic function $\Phi_t(a,b,x) = E\left[\exp(a\V_t+ibv_t)|v_0=x\right]$ given as following (Proposition 6.2.5 in \cite{LambertonLapeyre:1996})
\begin{equation}\label{eq:chfCIR}
\Phi_t(a,b,x) = A(t,a,b)\exp\left(B(t,a,b)x\right)
\end{equation}
with
\begin{align*}
A(t,a,b) &= \frac{
	\exp\left(\frac{\kappa^2\eta t}{\zeta^2}\right)
	}{
	\left(\cosh\left(\frac{\gamma t}{2}\right) 
	+\frac{\kappa-ib\zeta^2}{\gamma}\sinh\left(\frac{\gamma t}{2}\right)\right)
	^{2\kappa\eta/\zeta^2}}
\\
B(t,a,b) &= \frac{
	ib\left( \gamma\cosh\left(\frac{\gamma t}{2}\right)
		-\kappa\sinh\left(\frac{\gamma t}{2}\right)\right)
	+2ia\sinh\left(\frac{\gamma t}{2}\right)
	}{
	\gamma\cosh\left(\frac{\gamma t}{2}\right)
		+\left(\kappa-ib\zeta^2\right)\sinh\left(\frac{\gamma t}{2}\right)}
\\
\gamma &= \sqrt{\kappa^2-2\zeta^2ia}.
\end{align*}

\subsection{CGMY Process}
For $\alpha\in(0,2)$, $C, \lambda_+, \lambda_->0$, and $\mu\in\R$, an infinitely divisible random variable $X$ with characteristic function (Ch.F) 
\begin{align*}
&\phi_\textup{CGMY}(u; \alpha, C, \lambda_+, \lambda_-, \mu)= \phi_X(u) = E[e^{iuX}] \\
&=\exp\left((\mu-C\Gamma(1-\alpha)(\lambda_+^{\alpha-1}-\lambda_-^{\alpha-1})) iu
-C\Gamma(-\alpha)\left((\lambda_+-iu)^{\alpha}-\lambda_+^{\alpha}+(\lambda_-+iu)^{\alpha}-\lambda_-^{\alpha}\right)\right)
\end{align*} 
is referred to as the CGMY distributed random variable with parameters $(\alpha$, $C$, $\lambda_+$, $\lambda_-$, $\mu)$\footnote{The class of tempered stable processes has been
introduced under different names including: ``truncated \levy~flight” (\cite{Koponen:1995}), ``KoBoL” process (\cite{Boyarchenko_Levendorskii:2000}), ``CGMY” process
(\cite{CGMY:2002}), and classical tempered stable process (\cite{RachevKimBianchiFabozzi:2011a}). \cite{Rosinski:2007} and \cite{Bianchi_et_al:2010} generalized the notion of tempered stable processes.}. In this case, we denote $X\sim \textup{CGMY}(\alpha$, $C$, $\lambda_+$, $\lambda_-$, $\mu)$.

Let $C=(\Gamma(2-\alpha)(\lambda_+^{\alpha-2}+\lambda_-^{\alpha-2}))^{-1}$ and $\mu=0$. Then a CGMY random variable $Z\sim \textup{CGMY}(\alpha$, $C$, $\lambda_+$, $\lambda_-$, $\mu)$ has zero mean ($E[Z]=0$) and unit variance ($\var(Z)=1$). In this case, we say that $Z$ is standard CGMY distributed and denote $Z\sim \textup{stdCGMY}(\alpha$,  $\lambda_+$, $\lambda_-)$. Moreover, the Ch.F of $Z$ is given by
\begin{align}
\nonumber &
\phi_\textup{stdCGMY}(u; \alpha, \lambda_+, \lambda_-)  = \phi_Z(u) = E[e^{iuZ}] \\
\label{eq:chf_stdCTS} &
=\exp\left(\frac{\lambda_+^{\alpha-1}-\lambda_-^{\alpha-1}}{(\alpha-1)(\lambda_+^{\alpha-2}+\lambda_-^{\alpha-2})}iu
+\frac{
   (\lambda_+-iu)^{\alpha}-\lambda_+^{\alpha}+(\lambda_-+iu)^{\alpha}-\lambda_-^{\alpha}
   }
   {\alpha(\alpha-1)(\lambda_+^{\alpha-2}+\lambda_-^{\alpha-2})
   } 
   \right).
\end{align}

Since the CGMY distribution is purely non-Gaussian infinitely divisible, it generate a pure jump \levy~process $\{X_t\}_{t\ge0}$ such that $X_1$ $\sim$ $\textup{CGMY}(\alpha$, $C$, $\lambda_+$, $\lambda_-$, $\mu)$. In this case, we say that $(X_t)_{t\ge0}$ is CGMY process with parameters $(\alpha$, $C$, $\lambda_+$, $\lambda_-$, $\mu)$. The characteristic function (Ch.F) of $X_t$ is 
\begin{align*}
&\phi_{X_t}(u)=\exp(t\log(\phi_\textup{CGMY}(u;\alpha, C, \lambda_+, \lambda_-, \mu))).
\end{align*}
With the same argument, a pure jump~\levy process $\{Z_t\}_{t\ge0}$ such that $Z_1$ $\sim$ $\textup{stdCGMY}(\alpha$, $\lambda_+$, $\lambda_-)$ is referred to as the \textit{standard CGMY process} with parameters $(\alpha$, $\lambda_+$, $\lambda_-)$. The CGMY and standard CGMY processes are characterized by their \levy~symbols 
\[
\psi_\textup{CGMY}(u; \alpha, C, \lambda_+, \lambda_-, \mu) = \log \phi_\textup{CGMY}(u;\alpha, C, \lambda_+, \lambda_-, \mu)
\]
and 
\[
\psi_\textup{stdCGMY}(u; \alpha, \lambda_+, \lambda_-) = \log \phi_\textup{stdCGMY}(u;\alpha, \lambda_+, \lambda_-),
\]
respectively.

\subsection{\label{sec series representation}Series Representation of the CGMY Process}

 \cite{Rosinski:2007} introduced the series representation form for the tempered stable random variable and process, and it can be used for the CGMY sample path generation.
Assume that $X\sim \textup{CGMY}(\alpha$, $C$, $\lambda_+$, $\lambda_-$, $0)$. 
  Let $\{U_j\}_{j = 1,2,\cdots}$ be an independent and identically distributed (i.i.d.) sequence of uniform random variables on $(0,1)$. Let $\{E_j\}_{j = 1,2, \cdots}$ be i.i.d. sequences of exponential random variables with parameters 1, and let $\{\Gamma_j\}_{j = 1,2,\cdots}$ be a Poisson point process with parameter 1. Let $\{V_j\}_{j = 1,2, \cdots}$ be an i.i.d. sequence of random
variables in $\{\lambda_+, \lambda_-\}$ with $P(V_j=\lambda_+)=P(V_j=\lambda_-)=1/2$.
Suppose that $\{U_j\}_{j = 1,2, \cdots}$, $\{V_j\}_{j = 1,2, \cdots}$, $\{E_j\}_{j = 1,2, \cdots}$, and $\{\Gamma_j\}_{j = 1,2, \cdots}$ are independent. 
Then $X$ represented by the following series form:
\[
X = \sum_{j=1}^\infty\left[\left(\frac{\alpha \Gamma_j}{2 C}\right)^{-1/\alpha}\wedge E_j U_j^{1/\alpha}|V_j|^{-1}\right]\frac{V_j}{|V_j|}
+b,
\]
where $b = -C\Gamma(1-\alpha)\left(\lambda_+^{\alpha-1}-\lambda_-^{\alpha-1}\right)$.
Let $\{\tau_j\}_{j = 1,2, \cdots}$ be an i.i.d. sequence of uniform random variables on $(0,T)$ independent of $\{U_j\}_{j = 1,2, \cdots}$, $\{V_j\}_{j = 1,2, \cdots}$, $\{E_j\}_{j = 1,2, \cdots}$, and $\{\Gamma_j\}_{j = 1,2, \cdots}$. Suppose
\[
X_{t} = \sum_{j=1}^\infty\left[\left(\frac{\alpha \Gamma_j}{2 CT}\right)^{-1/\alpha}\wedge E_j U_j^{1/\alpha}|V_j|^{-1}\right]\frac{V_j}{|V_j|}1_{\tau_j\le t}
+tb_T, ~~~ t\in[0,T],
\]
where $b_T = 
-C\Gamma(1-\alpha)\left(\lambda_+^{\alpha-1}-\lambda_-^{\alpha-1}\right)$.
Then the the process $\{X_t\}_{t\in[0,T]}$ is the CGMY process with parameters  $(\alpha$, $C$, $\lambda_+$, $\lambda_-$, $0)$ for the time horizon $T>0$.

\section{Stochastic volatility version of the CGMY process}
Suppose $\{Z_t\}_{t\ge0}$ is the standard CGMY process with parameters $(\alpha$, $\lambda_+$, $\lambda_-)$ and $\{v_t\}$ is the stochastic volatility process given by CIR model in\eqref{eq CIR Process}. We define a process $\{L_t\}_{t\ge0}$ by
\begin{equation}\label{eq:svLevyPross} 
L_t = Z_{\V_t} + \rho v_t
\end{equation} 
where $\V_t = \int_0^t v_s ds$, and $\{Z_t\}_{t\ge0}$ is independent of the process $\{v_t\}_{t\ge0}$. 
The process $\{L_t\}_{t\ge0}$ is referred to as the \emph{stochastic volatility version of the CGMY process} or simply \emph{CGMYSV} process\footnote{
In \cite{Carr_et_al:2003}, $\{Z_t\}_{t\ge0}$ is assumed to a CGMY process, but we assume a standard CGMY process in this paper to simplify the model. The stochastic volatility \levy~process model is not a \levy~process in general.} with parameters $(\alpha$, $\lambda_+$, $\lambda_-$,  $\kappa$, $\eta$, $\zeta$, $\rho$, $v_0)$.
By \eqref{eq:chfCIR}, we obtain the characteristic function of $L_t$ as
\begin{equation}\label{eq phi Xt svlevy}
\phi_{L_t}(u) = \Phi_t(-i\psi_\textup{stdCGMY}(u; \alpha, \lambda_+, \lambda_-),\rho u,v_0),
\end{equation}
where $\phi_\textup{stdCGMY}(u; \alpha, \lambda_+, \lambda_-)$ is the characteristic function of $Z_1$ defined in \eqref{eq:chf_stdCTS}.

\begin{algorithm}
\SetAlgoLined
\KwResult{SVMYSV sample path}
 Let $T$ be the time horizon \;
 Let $M$, $J$, and $N$ be large positive integer \;
	$\varDelta t=T/M$,
	$v_{n,0} = v_0$,
	$c = \frac{2\kappa}{(1-e^{-\kappa \dt})\zeta^2}$,
	$C = (\Gamma(2-\alpha)(\lambda_+^{\alpha-2}+\lambda_-^{\alpha-2}))^{-1}$ \;
	$n = 1$ \;
 \While{$n\le N$}{
    $m = 1$\;
   \While{$m\le M$}{
   	$\xi = $ non-central $\chi^2$-distributed random variable with degrees of freedom $4\kappa\eta/\zeta^2$ and noncentrality parameter $2cv_{n, m-1}e^{-\kappa \dt}$ \;
   	$v_{n, m}=\xi/(2c)$ \;
   	$m = m+1$
   	}
 	$j = 1$, 	$\Gamma_0 = 0$\;
	 \While{$j\le N$}{
    	$U_j = $ uniform random numbers on $(0,1)$ \;
	   	$U'_j = $ uniform random numbers on $(0,1)$ \;
		$E_j = $ exponential random numbers with parameter 1 \;
		$E'_j = $ exponential random numbers with parameter 1 \;
		$\Gamma_j = \Gamma_{j-1}+E_j'$ \;
		\eIf{$U'_j \le 0.5 $}{
			$V_j = \lambda_+$
		   }{
			$V_j = \lambda_-$
		  }
	 $\tau_j = $ uniform random numbers on $(0,T)$ \;
   	$c(\tau_j) = C\sum_{k=1}^M v_{n, k-1} 1_{(k-1)\dt < \tau_j \le k\dt}$ \;
	 $j = j+1$ \;
	 }
   $m = 1$,   $Y_{n,0} = 0$ \;
   \While{$m\le M$}{
   	$b_m = -\frac{v_{n,m-1}\left(\lambda_+^{\alpha-1}-\lambda_-^{\alpha-1}\right)}{(1-\alpha)(\lambda_+^{\alpha-2}+\lambda_-^{\alpha-2})}$\;
   	$
Y_{n, m} = Y_{n, m-1} + \sum_{j=1}^J\left[\left(\frac{\alpha \Gamma_j}{2 c(\tau_j)T}\right)^{-1/\alpha}\wedge E_j U_j^{1/\alpha}|V_j|^{-1}\right]\frac{V_j}{|V_j|}1_{(m-1)\dt <\tau_j\le m\dt}
+ b_m\dt \;
$
	$L_{n,m} = Y_{n,m}+\rho v_{n,m}$ \;
	$m = m+1$\;
	}
	$n = n+1$\;
 }
 \caption{\label{Algorithm} CGMYSV sample path generation}
\end{algorithm}

\subsection{Series representation of the CGMYSV Process}
Suppose that we have a CIR process $\{v_t\}_{t\ge0}$ with parameters $\kappa, \eta,$ and $\zeta$ as defined in \eqref{eq CIR Process}, $\V_t = \int_0^t v_sds$ for $t>0$, and suppose that $\{\mathcal F^v_t\}_{t\ge0}$ is natural filtration generated by  $\{v_t\}_{t\ge0}$.
Let $P=\{0= t_0<t_1<\cdots<t_m<\cdots < M=T\}$ be the partition of time, $\dt_m = t_m - t_{m-1}$ for $m \in \{1, 2,\cdots, M\}$, and let $||P|| = \max\{\dt_m\,|\,m = 1,2,\cdots, M\}$. 
Suppose that $\{v_{t_m}\}_{t_m\in P}$ and $\{L_{t_m}\}_{t_m\in P}$ are discrete sub-sequences of the CIR process and the CGMYSV process, respectively. Let $\varDelta L_{t_m} =  L_{t_m}-L_{t_{m-1}}$, $\varDelta \V_{t_m} = \V_{t_m}-\V_{t_{m-1}}$, and $\varDelta v_{t_m} = v_{t_m}-v_{t_{m-1}}$.
Then we have
\[
\varDelta L_{t_m} |_{\mathcal F^v_{t_{m-1}}} \disteq (Z_{\V_{t_m}-\V_{t_{m-1}}})|_{\mathcal F^v_{t_{m-1}}} + \rho(v_{t_m}-v_{t_{m-1}})|_{\mathcal F^v_{t_{m-1}}} = Z_{\varDelta \V_{t_m}}|_{\mathcal F^v_{t_{m-1}}} + \rho(\varDelta v_{t_m}|_{\mathcal F^v_{t_{m-1}}}),
\]
where
\[
Z_{\varDelta \V_{t_m}}|_{\mathcal F^v_{t_{m-1}}} \sim \textup{CGMY}\left(\alpha, C(\varDelta \V_{t_m}|_{\mathcal F^v_{t_{m-1}}}), \lambda_+, \lambda_-, 0\right),
\]
and $C = \left(\Gamma(2-\alpha)(\lambda_+^{\alpha-2}+\lambda_-^{\alpha-2})\right)^{-1}$.
Since we approximate 
\[
\varDelta \V_{t_m}|_{\mathcal F^v_{t_{m-1}}} = \int_{t_{m-1}}^{t_m} v_t \,dt \approx v_{t_{m-1}}\dt_m,
\]
we have
\[
Z_{\varDelta \V_{t_m}}|_{\mathcal F^v_{t_{m-1}}} \approx Z_{v_{t_{m-1}}\dt_m}|_{\mathcal F^v_{t_{m-1}}} \sim \textup{CGMY}\left(\alpha, C v_{t_{m-1}} \dt_m, \lambda_+, \lambda_-, 0\right).
\]
Suppose $\{Y_{t_m}\}_{t_m\in P}$ is a process defined by 
 \[
 Y_{t_m} 
 = Y_{t_{m-1}} +Z_{v_{t_{m-1}}\dt_m}|_{\mathcal F^v_{t_{m-1}}}, ~~~ m = 1,2,\cdots, M
 \]
with $Y_0 = 0$, then $Y_{t_m}\approx Z_{\V_{t_m}}|_{\mathcal F^v_{t_{m-1}}}$ and $\{Y_{t_m}\}_{t_m\in P}$ is an approximation of the process $\{Z_{\V_{t_m}}|_{\mathcal F^v_{t_{m-1}}} \}_{t_m\in P}$.

By the series representation, we have
\[
Z_{v_{t_{m-1}}\dt_m}|_{\mathcal F^v_{t_{m-1}}} \disteq \sum_{j=1}^\infty\left[\left(\frac{\alpha \Gamma_j}{2 c_{m}\dt_m}\right)^{-1/\alpha}\wedge E_j U_j^{1/\alpha}|V_j|^{-1}\right]\frac{V_j}{|V_j|}+b_m\dt_m
\]
where
$b_m =  -c_{m}\Gamma(1-\alpha)\left(\lambda_+^{\alpha-1}-\lambda_-^{\alpha-1}\right)$, 
$c_{m}=v_{t_{m-1}} C$,
 and $\{U_j\}_{j = 1,2, \cdots}$, $\{V_j\}_{j = 1,2, \cdots}$, $\{E_j\}_{j = 1,2, \cdots}$, and $\{\Gamma_j\}_{j = 1,2, \cdots}$ are given in Section \ref{sec series representation}.
The same argument as the relation between series representation of the CGMY process presented in Section \ref{sec series representation}, we can define a series representation of $Y_{t_m}$ as follows:
\begin{equation}\label{Eq:2}
Y_{t_m} = \sum_{j=1}^\infty\left[\left(\frac{\alpha \Gamma_j}{2 c(\tau_j)T}\right)^{-1/\alpha}\wedge E_j U_j^{1/\alpha}|V_j|^{-1}\right]\frac{V_j}{|V_j|}1_{\tau_j\le t_m}
+ \sum_{k=1}^m b_k\dt_k
\end{equation}
where
\[
b_k = -\frac{v_{t_{k-1}}\left(\lambda_+^{\alpha-1}-\lambda_-^{\alpha-1}\right)}{(1-\alpha)(\lambda_+^{\alpha-2}+\lambda_-^{\alpha-2})},
\]
\[
c(\tau_j) = C\sum_{m=1}^M v_{t_{m-1}} 1_{t_{m-1} < \tau_j \le t_m},
\]
and  $\{\tau_j\}_{j = 1,2, \cdots}$ is an i.i.d. sequence of uniform random variables on $(0,T)$ independent of $\{U_j\}_{j = 1,2, \cdots}$, $\{V_j\}_{j = 1,2, \cdots}$, $\{E_j\}_{j = 1,2, \cdots}$, and $\{\Gamma_j\}_{j = 1,2, \cdots}$. Therefore, we have
\begin{equation}\label{Eq:3}
L_{t_m}|_{\mathcal F^v_{t_{m-1}}} = Z_{\V_{t_m}}|_{\mathcal F^v_{t_{m-1}}} + \rho v_{t_m}|_{\mathcal F^v_{t_{m-1}}} \approx Y_{t_m} + \rho v_{t_m}|_{\mathcal F^v_{t_{m-1}}}.
\end{equation}
Combining equations \eqref{Eq:2} and \eqref{Eq:3}, we can generate sample path of the CGMYSV process as Algorithm \ref{Algorithm}.


\subsection{Simulation of the CGMYSV Process}

\begin{figure}[t]
\includegraphics[width = 8cm]{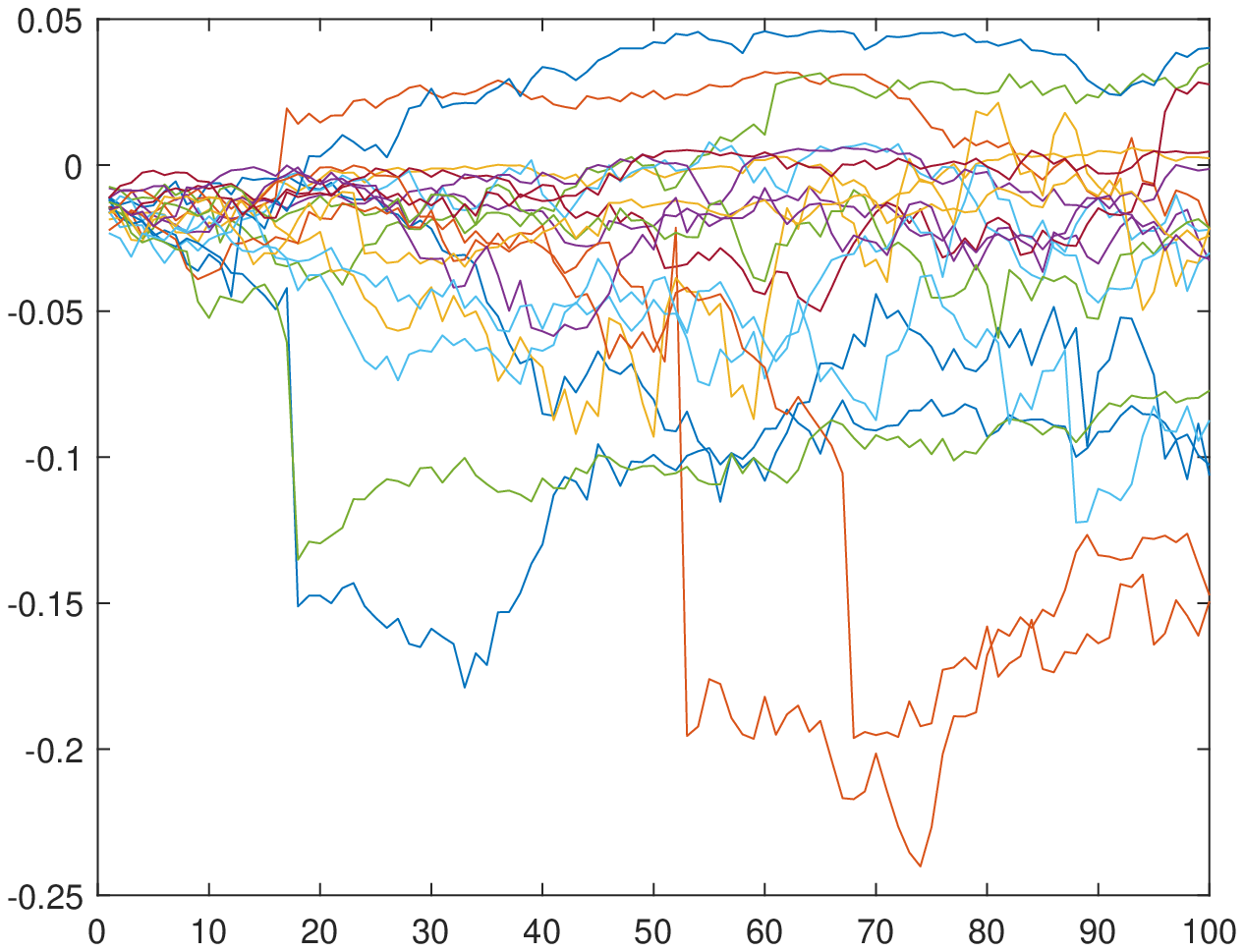}
\includegraphics[width = 8cm]{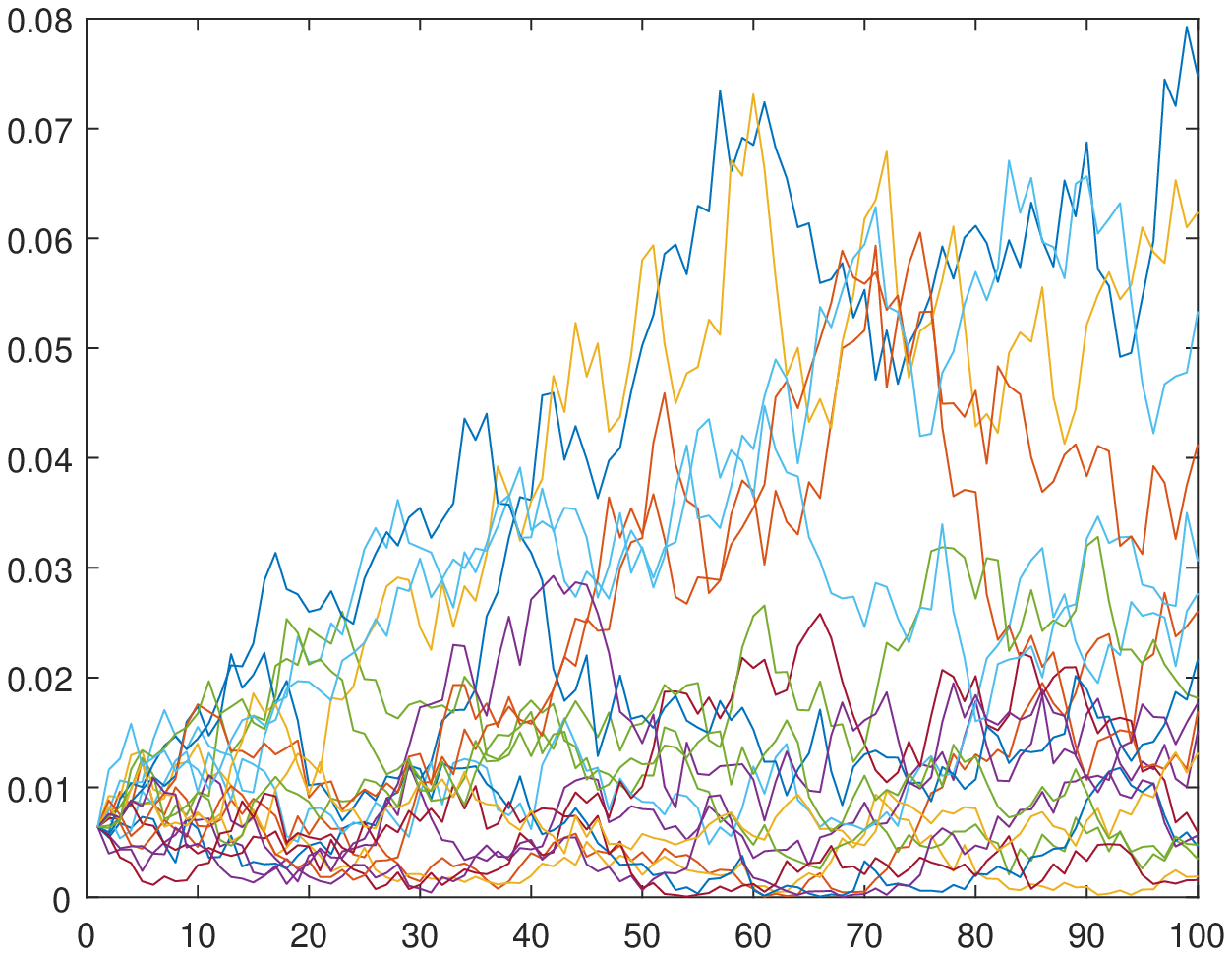}
\caption{\label{CGMYSV Simulation Samplepath}CGMYSV sample paths (left) and CIR sample paths (right).}
\end{figure}

\begin{figure}[t]
\includegraphics[width = 8cm]{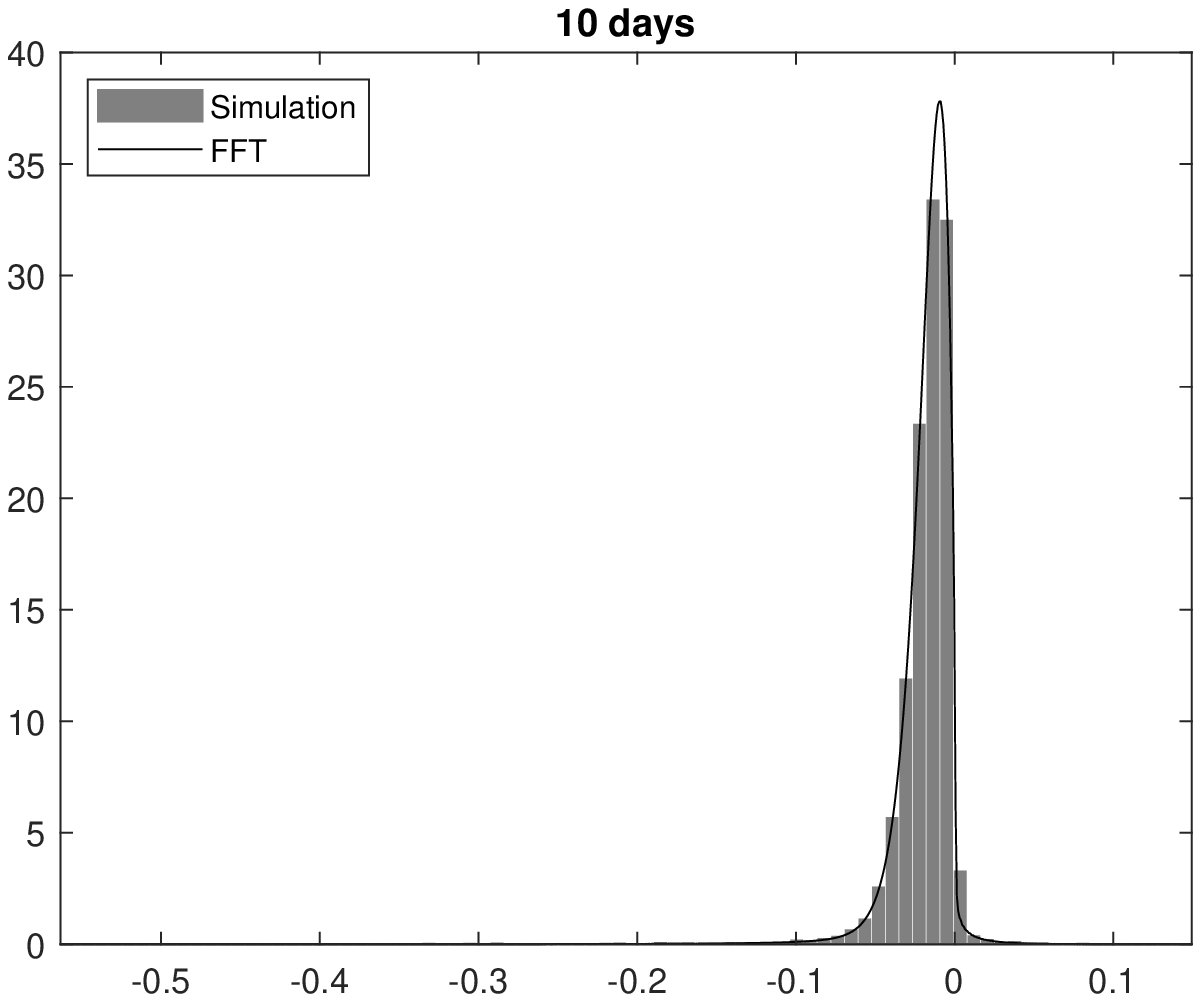}
\includegraphics[width = 8cm]{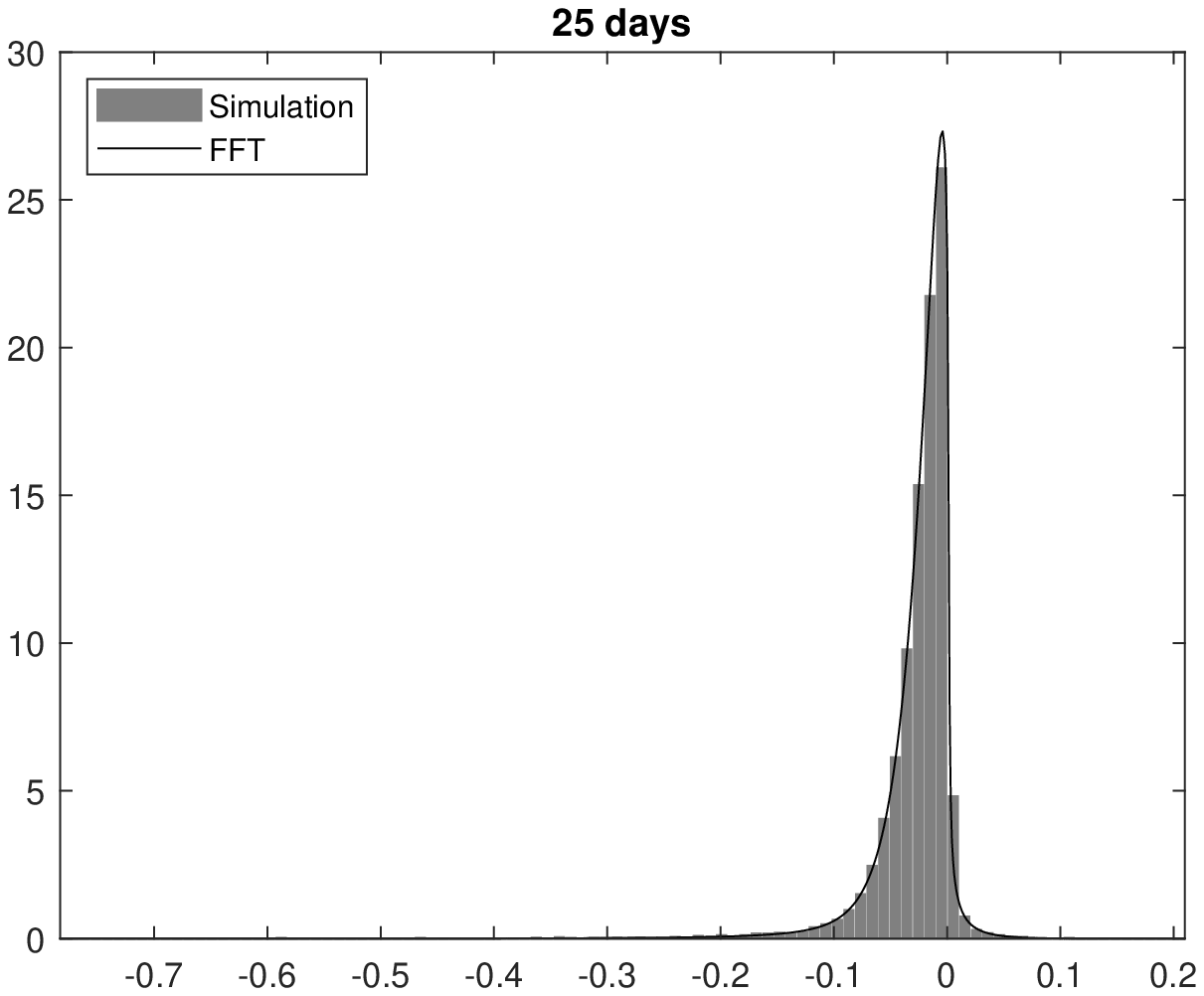}\\
\includegraphics[width = 8cm]{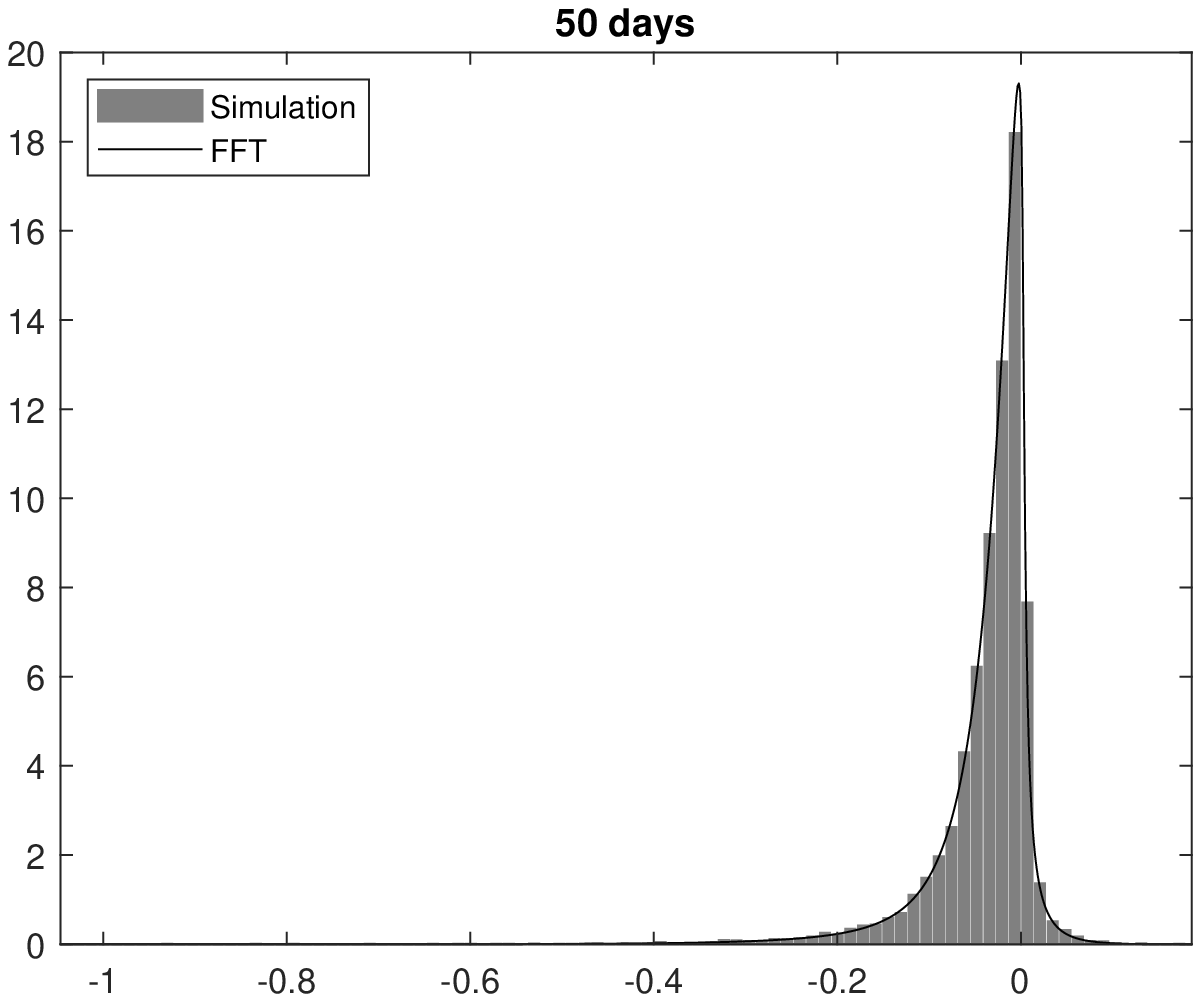}
\includegraphics[width = 8cm]{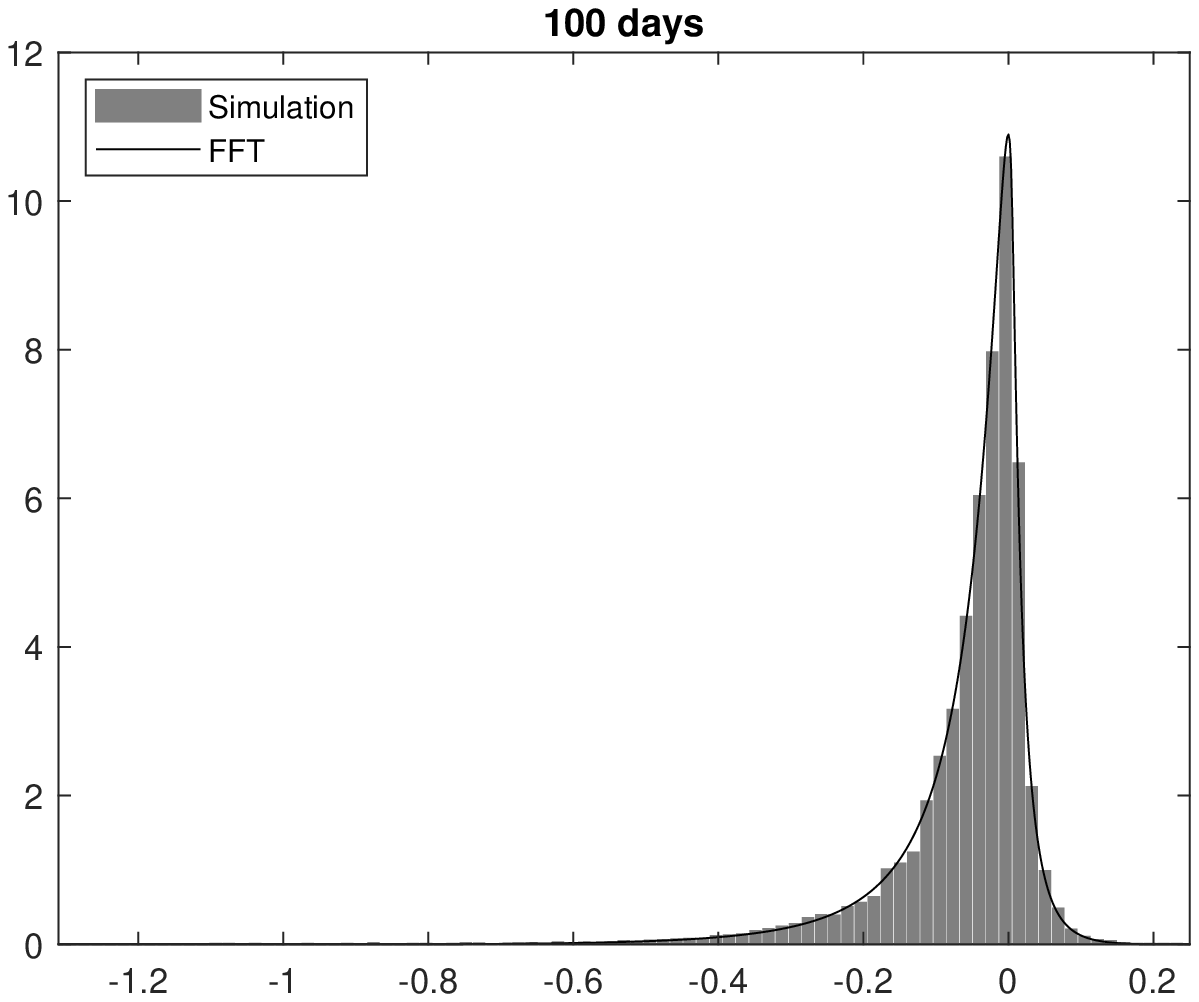}
\caption{\label{fig:Simulation PDF vs FFT pdf}CGMYSV pdfs based on the simulated sample path (gray bar-plots) vs pdf using FFT method (solid curves). Distributions of $X_t$ are for $t=10\varDelta t$ (top-left), $t=25\varDelta t$ (top-right), $t=50\varDelta t$ (bottom-left), and $t=10\varDelta t$ (bottom-right), where $\varDelta t = 1/252$ is one day of year fraction.}
\end{figure}

In order to verify the performance of Algorithm \ref{Algorithm}, we generate a set of example sample paths of the CGMYSV process $\{L_t\}_{t\ge0}$ with parameters
$\alpha = 0.52$, $\lambda_+ =25.46$, $\lambda_- = 4.604$, $\kappa = 1.003$, 
$\eta = 0.0711$, $\zeta= 0.3443$, $v_0 = 0.0064$, and $\rho = -2.0280$.
We set $M=100$, $J=1024$, $N=10,000$ , and $\dt = 1/252$ which is the annual fraction of 1 day.
Example 20 sample-paths are presented in the first plate of Figure \ref{CGMYSV Simulation Samplepath}. The second plate of the figure is for 20 sample path of CIR process. For goodness of fit test for the generated path, we perform Kolmogorov-Smirnov test. We compare the distribution of 10-days simulated random numbers $\{L_{n, 10}|n = 1,2,\cdots, N\}$ with the distribution of $L_{10\dt}$. The cumulative distribution function of $L_{10\dt}$ can be obtained by the Ch.F of the CGMYSV using the inverse Fourier-Transform methos (See \cite{CGMY:2002} and \cite{RachevKimBianchiFabozzi:2011a} more details). Table \ref{Table KS Test} presents the result of the KS test, and it has 70.29\% $p$-value and it is not rejected at the 5\% significant level. Using the same arguments, we perform KS test for 25-days, 50-days, and 100-days simulated random numbers. They are not rejected at the 5\% significant level, either. We graphically compare the empirical probability density function (pdf) of the simulated sample path and the CGMYSV pdfs for those four cases. We draw empirical pdfs using gray bar charts and draw solid lines for CGMYSV pdfs in four plates in Figure \ref{fig:Simulation PDF vs FFT pdf}.

\section{The CGMYSV Option Pricing Model}
In this section we discuss the option pricing model on the CGMYSV model. We define the model and calibrate parameters using European style S\&P 500 index option (SPX option) and American style S\&P 100 index option (OEX option).

Let $r$ and $q$ be the risk free rate of return and the continuous dividend rate of a
given underlying asset, respectively. The risk-neutral price process $\{S_t\}_{t\ge 0}$ of  a given underlying asset is assumed as
\begin{equation}\label{eq:rn cgmysv stock price process}
S_t = \frac{S_0 \exp((r-q)t+L_t)}{E\left[\exp(L_t)\right]}
\end{equation}
where $\{L_t\}_{t\ge 0}$ is the CGMYSV process with parameters $(\alpha$, $\lambda_+$, $\lambda_-$, $\kappa$, $\eta$, $\zeta$, $\rho$, $v_0)$. By \eqref{eq phi Xt svlevy}, we also have
\[
S_t = \frac{S_0 \exp((r-q)t+L_t)}{\Phi_t(-i\log\phi_{stdCTS}(-i; \alpha, \lambda_+, \lambda_-),-\rho i,v_0)}.
\]
\subsection{Calibration to European Options}
\begin{figure}[h]
\centering
\includegraphics[width = 14cm]{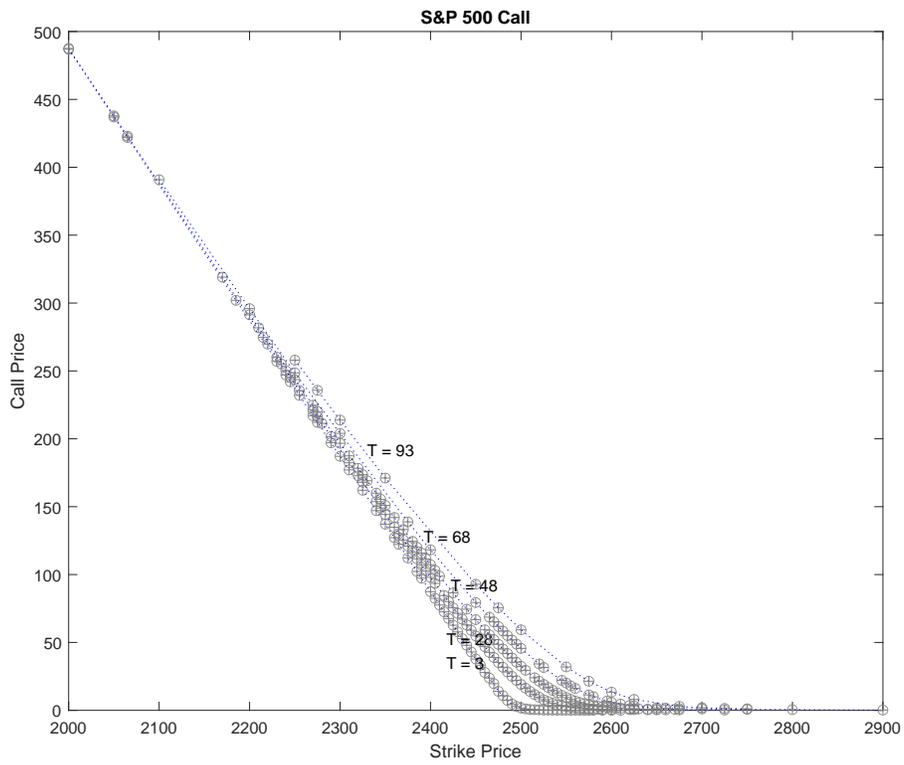}
\includegraphics[width = 14cm]{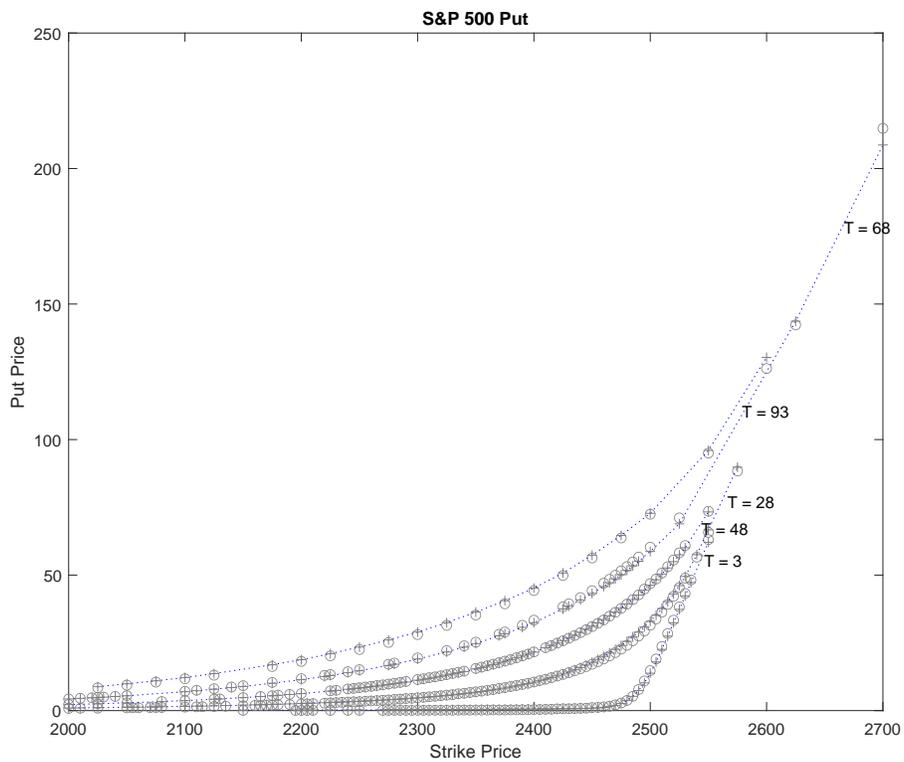}
\caption{\label{Figure:CalibrationEuropeanCallPut}Observed SPX option price and model prices calibrated to the market prices for Call (top) and put (bottom) on September 11, 2017.  `$\circ$' stands for the market price and `+' stands for the FFT price.}
\end{figure}

On the risk neutral price process $\{S_t\}_{t\ge 0}$ defined by \eqref{eq:rn cgmysv stock price process}, the European call and put prices with the strike price $K$ and the time to maturity $T$ are equal to $C(K,T) = e^{-rT} E[(S_t-K)^+]$ and $C(K,T) = e^{-rT} E[(K-S_t)^+]$, respectively. Moreover, the Fast-Fourier-Transform (FFT) method by \cite{CarrMadan:1999} and \cite{Boyarchenko_Levendorskii:2000}, we can calculate European call/put prices numerically. We calibrate the CGMYSV parameters $(\alpha$, $\lambda_+$, $\lambda_-$, $\kappa$, $\eta$, $\zeta$, $\rho$, $v_0)$ using the SPX option prices on September 11, 2017. We observed 247 call prices and 289 put prices on the day. The S\&P 500 price $S_0$, risk-free rate of return, and continuous dividend rate at the day were $S_0 = 2488.11$, $r = 1.213\%$, and $q = 1.884\%$ respectively.
The calibration results for SPX calls and puts are provided in Table \ref{Table:CalibrationEuropeanCallPut}. Figure \ref{Figure:CalibrationEuropeanCallPut} shows observed SPX call and put prices ( drawn by `$\circ$'), and calibrated CGMYSV prices using FFT (drawn by `$+$'). 

We recalculate the European call and put prices using Monte-Carlo Simulation (MCS) method with the calibrated parameters in Table \ref{Table:CalibrationEuropeanCallPut}. The sample paths of the MCS method are generated by Algorithm \ref{Algorithm}. The number of sample paths is 10,000 in this investigation.

To compare the MCS method with the FFT method, we use the four error estimators:the average absolute error (AAE), the average absolute error as a percentage of the mean price (APE), the average relative percentage error (ARPE), and the root mean square error (RMSE) (see \cite{schoutens2003lpf}).\footnote{The measures are computes as follows:
$ \textup{AAE} \!=\!  \sum_{j=1}^N \frac{|\widehat{P}_j -{P}_j|}{N}$, $
 \textup{APE} \!=\! \frac{\sum_{j=1}^N |\widehat{P}_j -{P}_j|/N}{\sum_{j=1}^N\widehat{P}_j/N}$, $
 \textup{ARPE} \!=\! \frac{1}{N}\sum_{j=1}^N \frac{|\widehat{P}_j - {P}_j|}{\widehat{P}_j}$, 
 \textup{ and }
 $\textup{RMSE} \!=\! \sqrt{\frac{1}{N}\sum_{j=1}^N \frac{(\widehat{P}_j -{P}_j)^2}{N}},
$
where $N$ is the number of observations, and $\widehat{P}_j$ and ${P}_j$ denote the model price and the observed market call/put prices, respectively. }
Those four error estimators for the FFT method and the MCS method are in Table \ref{Table:Errors MCS vs FFT}. Both call and put cases, the MCS method has larger error estimators than the FFT method. That is not surprising because we calibrate those parameters using the FFT method. The table says that the four error estimators of MCS method are similar to those of the FFT method. That means the sample path generation with Algorithm \ref{Algorithm} performs well, and prices by the MCS are similar performance as FFT method.

\begin{figure}[h]
\centering
\includegraphics[width = 14cm]{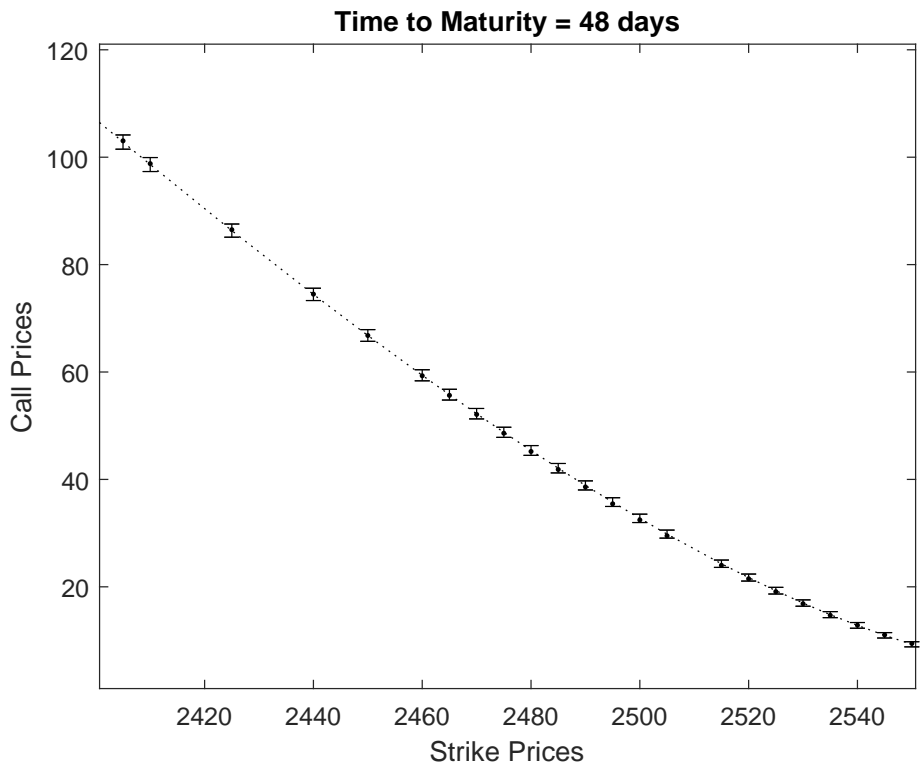}
\includegraphics[width = 14cm]{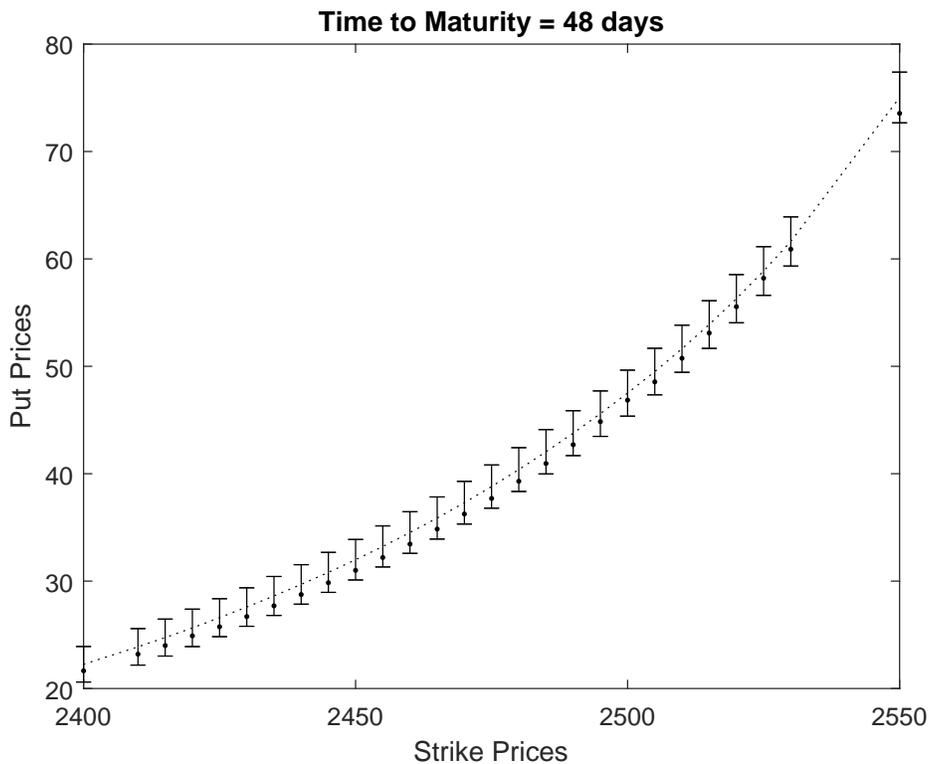}
\caption{\label{Figure:MCSEuropeanCallPutCI}Confidence Intervals for the MCS option prices. Dots are observed market prices, dot-coves are MCS prices, and `I' shape bars are 95\% confidence intervals of MCS prices. The first (top) plate is for call option pricing and the second (bottom) plate is for put option pricing.  }
\end{figure}

In this option pricing with MCS method, we also obtain standard error for each 247 call and 289 put options, but we do not provide them all because of the space limitation. Instead, we show MCS prices with the 95\% confidence interval in Figure \ref{Figure:MCSEuropeanCallPutCI} only for the case $2,400<K<2,600$ and time to maturity 48 days.  

\begin{figure}
\includegraphics[width = 14cm]{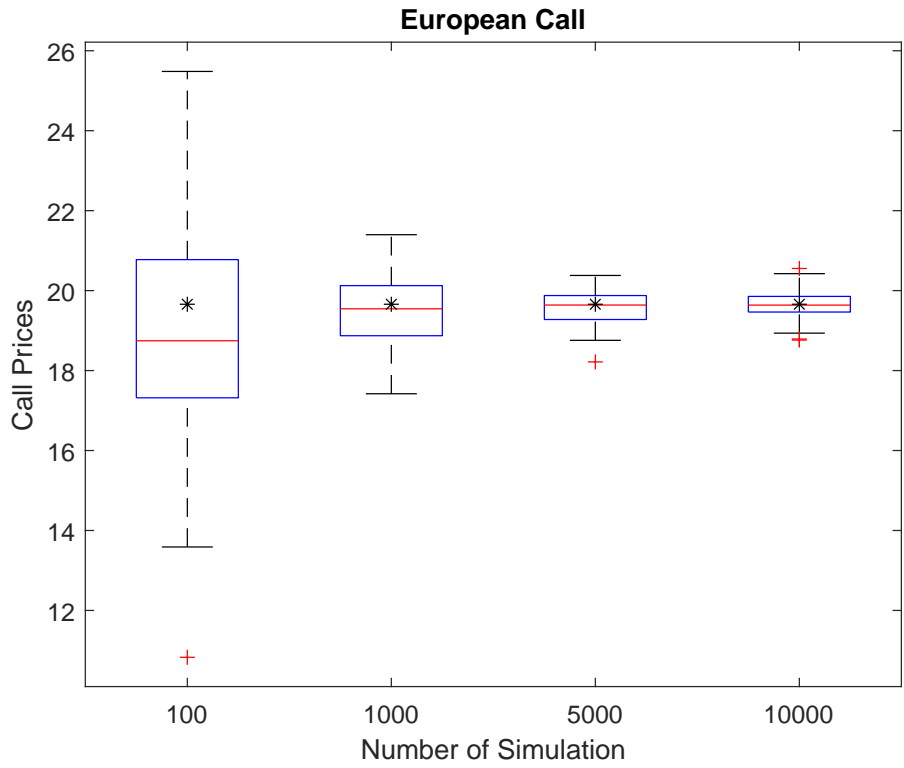}\\
\includegraphics[width = 14cm]{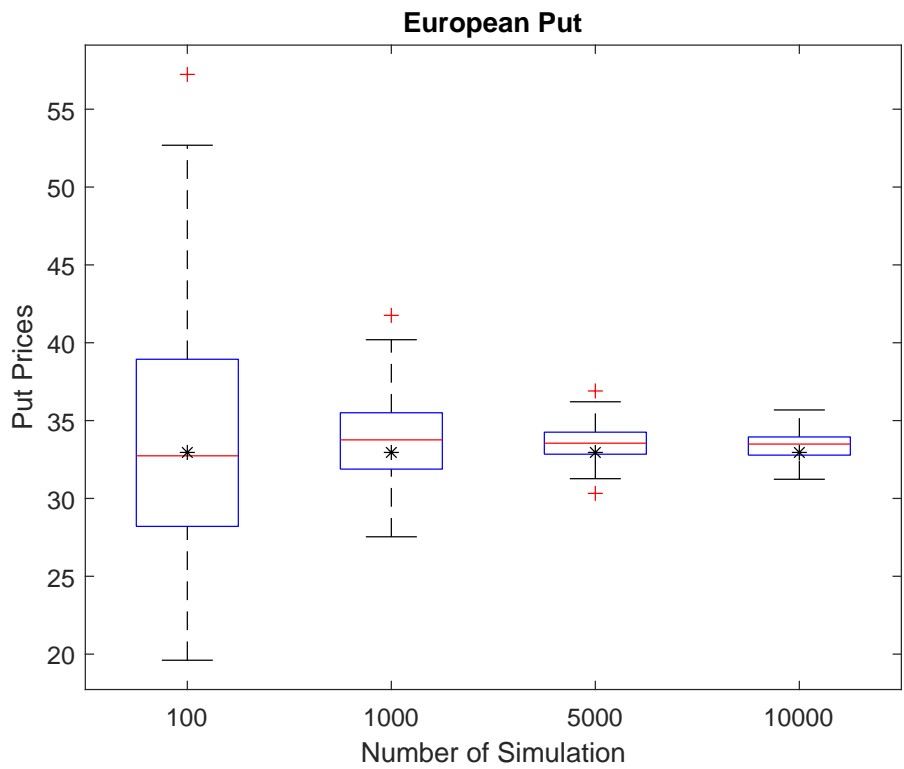}
\caption{\label{Figure:BootStrapping}Boot strapping for call (top) and put (bottom).  }
\end{figure}

Finally, we perform the bootstrapping. We select an at-the-money call and an at-the-money put of $K = 2,500$ and $T= 28\, days$ as an example, and calculate call and put prices with MCS parameters in Table \ref{Table:CalibrationEuropeanCallPut}, respectively. Table \ref{Table:ATM} shows that the MCS prices and their standard errors for 100, 1,000, 5,000, and 10,000 number of sample paths. We repeat this process 100 times and draw boxplots. Boxplots for call and put for each number of sample paths are the up plate and the bottom plate of Figure \ref{Figure:BootStrapping}. Stars in those boxplots are the call/put prices using FFT method. We can observe that the number of sample paths increases, then the MCS prices close to the FFT price and dispersions are reduced.

\subsection{Calibration to American Options}
We see that the sample path generation method using the series representation works for MCS of European option pricing in the previous section. In this section, we discuss the American option pricing with the same sample path generation method.
We use Least Square Regression Method (LSM) by \cite{Longstaff_Schwartz:2001} for American option pricing with MCS. When we do the regression for the expected value of option, we use $S_t$, $S_t^2$, $\sigma_t$, $\sigma_t^2$ and $\sigma_t S_t$ as independent variables, following the idea in Chapter 15 of \cite{RachevKimBianchiFabozzi:2011a}.

For empirical illustration, we use market prices of the OEX option, which is American style. We calibrate parameters of the CGMYSV model with fixed seed numbers for each random number generation. That is, we fix a seed number of $\chi^2$ random number generator in the CIR process, and we generate uniform and exponential random numbers $U_j$, $U'_j$, $E_j$, $E'_j$ and $\tau_j$ with predefined seed numbers, and fix them. Then we set the model parameters, generate sample paths using  Algorithm \ref{Algorithm} with the fixed seed number and the fixed random number sets, and then calculate American option price using LSM. Repeat that process and find the optimal parameters to minimize RMSE. As a benchmark, we calibrate the parameters of the CGMY option pricing model (See \cite{CGMY:2002}) to the OEX option prices using LSM with sample path generated by the series representation explained in Section \ref{sec series representation}.

The calibration results are presented in Table \ref{Table:CalibrationAmericanOption}.
We calibrate the CGMY and the CGMYSV model parameters for 12 Wednesdays in 2015 and 2016 exhibited in the table. The four error estimators, AAE, APE, ARPE, and RMSE, are also provided in Table \ref{Table:4errors}. Since the smaller error estimator means the better calibration performance, smaller errors are written in bold letters for each day. This table shows that the CGMYSV calibration performs better than CGMY calibration except for the cases of February 10, 2016 and June 10, 2015. On March 9, 2016, AAE and APE of the CGMY model are less than those of CGMYSV, but ARPE and RMSE of CGMY are larger than those of CGMYSV. ARPE of CGMY is smaller than that of CGMYSV on November 10, 2015, but the other three error values of CGMY are larger than CGMYSV. Therefore, we can conclude that the CGMYSV option pricing model performs typically better than the CGMY option pricing model, except in a few cases in this investigation. Hence, LSM with Algorithm \ref{Algorithm} works well in the American option calibration.

\begin{figure}
\centering
\includegraphics[width = \textwidth]{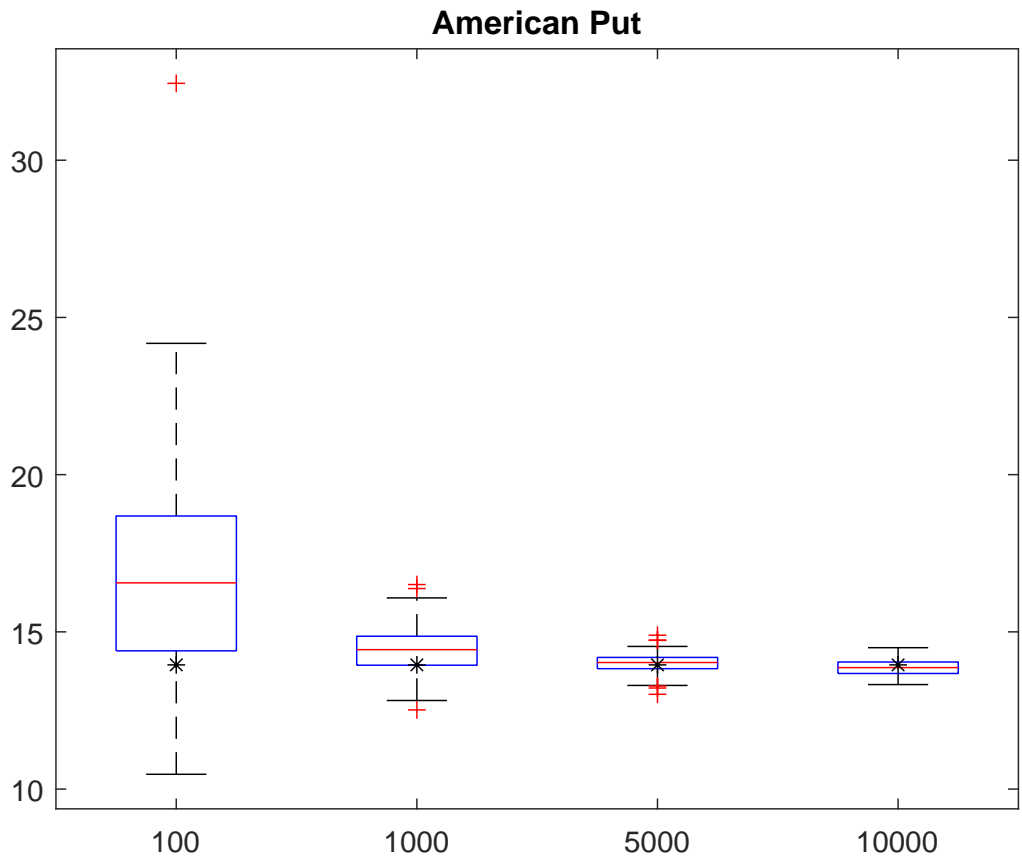}
\caption{\label{Figure:BootStrappingOEXPut}Boot strapping for OEX put option with time to maturity of $T = 31 days$ and strike price $K = 910$.  }
\end{figure}

\begin{sidewaysfigure}
\centering
\includegraphics[width = \textwidth]{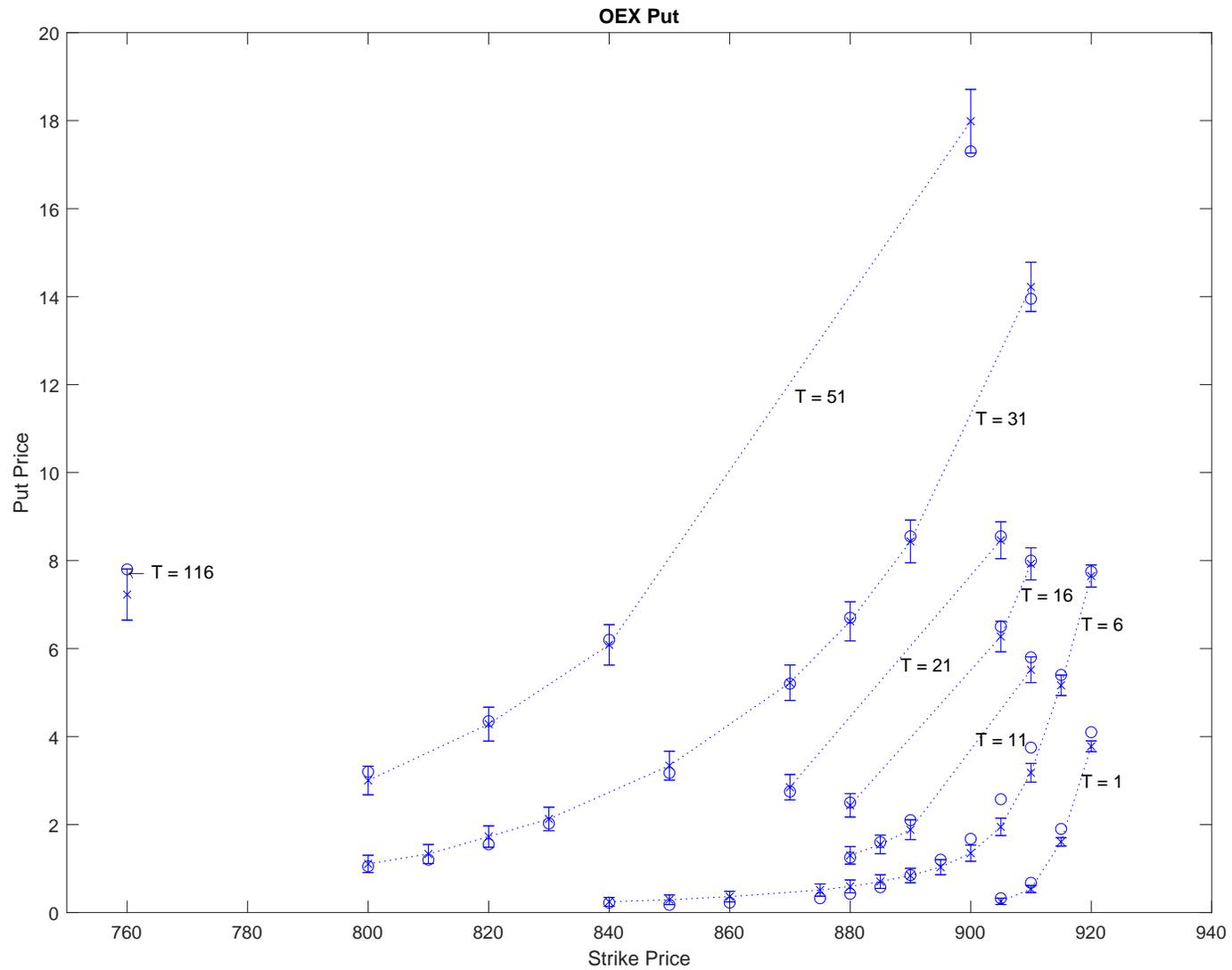}
\caption{\label{Fig:AmericanPut20160406}
OEX put prices on April 6, 2016 and LSM prices with their confidence intervals. Circles (`$\circ$') are observed OEX put prices, `$\times$' points are MCS prices, and `I' shape bars are 95\% confidence intervals of LMS prices. }
\end{sidewaysfigure}

Finally, we perform the bootstrapping. We selected the at-the-money put for the strike price $K = 910$ and the days to maturity $T= 31$ days on April 6, 2016. Put prices are obtained by LSM using parameters calibrated to the day provided in Table \ref{Table:CalibrationAmericanOption}. On the day, the underlying S\&P 100 index price was 918.21, and the market put price was 13.95 for the strike price 910 and 31 days to maturity. Table \ref{Table:OexMCSExample} shows that the LSM prices and their standard errors for 100, 1,000, 5,000, and 10,000 number of sample paths. The LSM prices approach to the market price and the standard error decreases as the number of sample paths increases. We repeat this process 100 times and present boxplots for those 100 prices, as Figure \ref{Figure:BootStrappingOEXPut}. Stars in those boxplots are the market put prices. We can observe that the LSM prices close to the market price and dispersions are reduced as the number of sample paths increases.
Additionally, Figure \ref{Fig:AmericanPut20160406} provides a graphical illustration of the calibration for April 6, 2016. Calibrated CGMYSV prices are drawn by `$\times$', the market observed prices are drawn by `$\circ$', and the 95\% confidence intervals are marked by 'I' shape. The day to maturities $T$ are written on the plate.

\subsection{Asian and Barrier options}
The sample path generation method for the CGMYSV model can be used for Asian and Barrier option pricing. In this section, we briefly show examples of Asian and Barrier option pricing using MCS with the sample path generated by Algorithm \ref{Algorithm}.

We generate 10,000 sample path of the CGMYSV model with parameters 
$\alpha = 0.52$, $\lambda_+ =25.46$, $\lambda_- = 4.604$, $\kappa = 1.003$, 
$\eta = 0.0711$, $\zeta= 0.3443$, $v_0 = 0.0064$, and $\rho = -2.0280$.
Then we generate the underlying price process $\{S_t\}_{t\ge0}$ using \eqref{eq:rn cgmysv stock price process} where $S_0=2,488$, $r = 0.0121$, $d = 0.0188$. 

\begin{figure}
\centering
\includegraphics[width = 12cm]{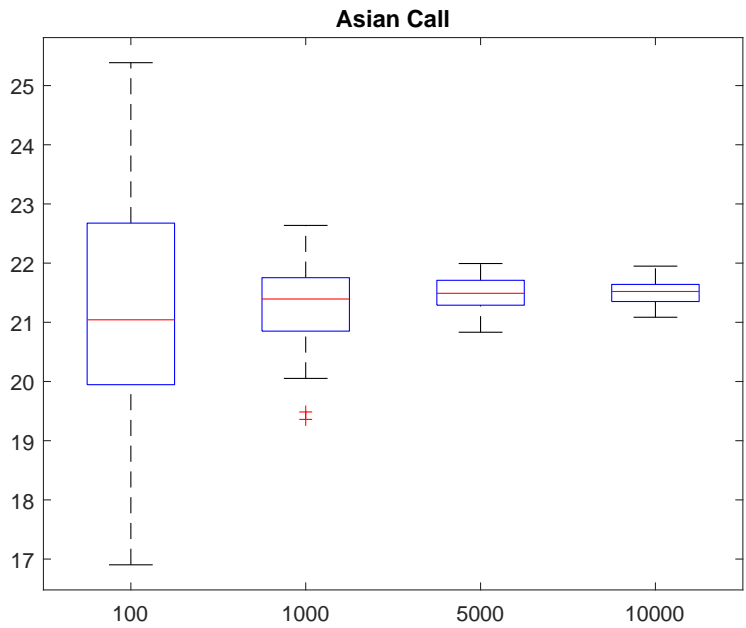}
\includegraphics[width = 12cm]{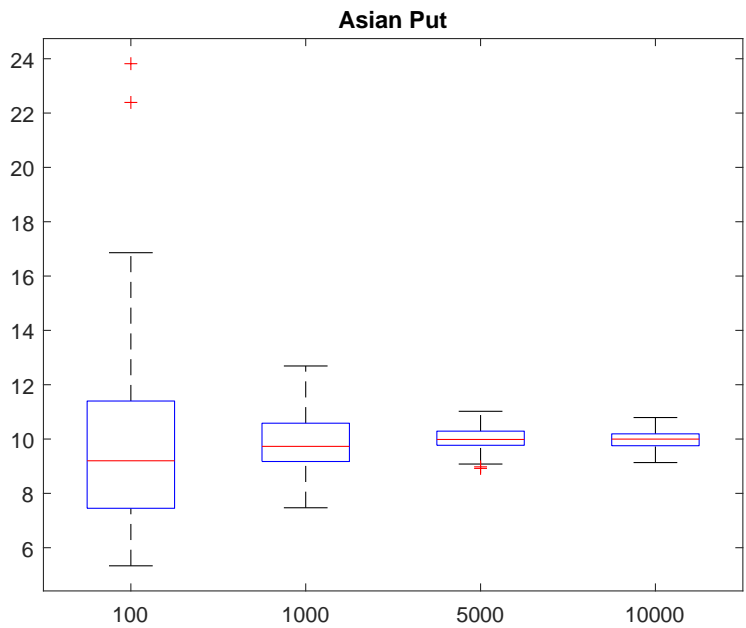}
\caption{\label{Fig:BootStrappingAsian}Boot strapping for Asian call (top) \& put (bottom).}
\end{figure}

For the Asian option, we consider the arithmetic average call and put where the strike price $K=2,500$ and the time to maturity $T = 25 days$. Table \ref{Table:Asian} shows that the MCS prices for Asian call \& put and their standard errors for 100, 1,000, 5,000, and 10,000 number of sample paths.
The the standard error of the MCS prices decreases as the number of sample paths increases. 
We repeat this process 100 times and present boxplots for those 100 prices, as Figure \ref{Fig:BootStrappingAsian}. We can observe that the MCS prices converge, and dispersions are reduced as the sample paths increase.

\begin{figure}
\centering
\includegraphics[width = 12cm]{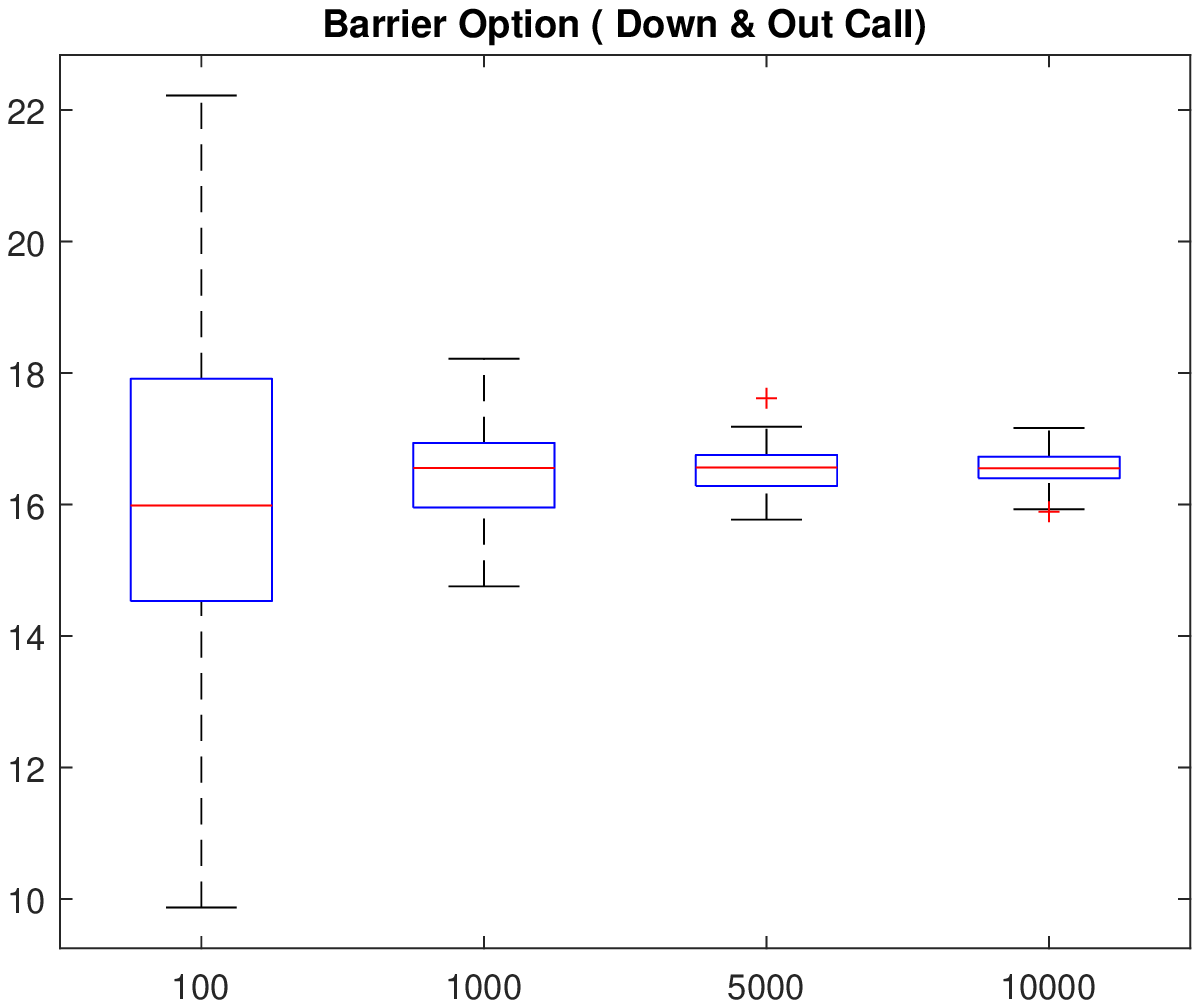}
\includegraphics[width = 12cm]{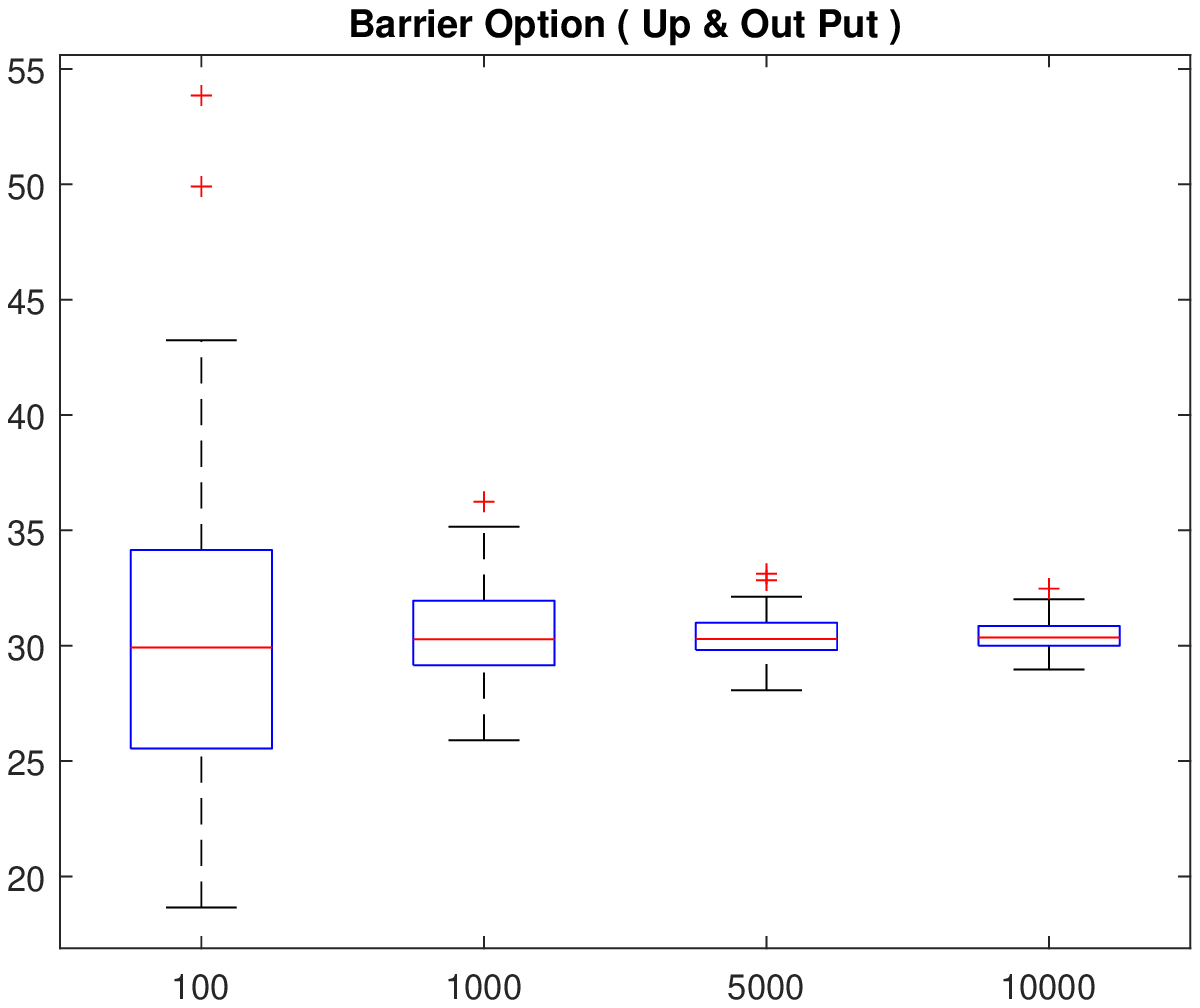}
\caption{\label{Fig:BootStrappingBarrier}Boot strapping for Barrier options: the down-and-out call (top) and the up-and-out put (bottom).}
\end{figure}

With the same argument, we find MCS price for the Barrier options. We consider the down-and-out call and the up-and-out put Barrier options with the strike price $K=2,500$ and the time to maturity $T = 25 days$. Barrier of the down-and-out call and the up-and-out put are $2,400$ and $2,750$, respectively. Table \ref{Table:Barrier} shows that the MCS prices and their standard errors for 100, 1,000, 5,000, and 10,000 number of sample paths. The standard error of the MCS prices decreases as the number of sample paths increases. 
We repeat this process 100 times and present boxplots for those 100 prices, as Figure \ref{Fig:BootStrappingBarrier}. We can also observe that the MCS prices converge, and dispersions are reduced as the sample paths increase.

\section{Conclusion}
In this paper, we develop the CGMYSV sample path generation algorithm using the series representation. The series representation method's performance is tested by comparing the simulated distribution to the pdf calculated by the inverse Fourier transform method. We apply the sample path generation method to European and American option pricing with MCS and LSM. We compare the MCS method to the FFT method in European option pricing with SPX option market data. Also, we calibrate the parameters of the CGMYSV model to the American style OEX option using LSM. We measure the performance of the calibration using four error estimators and the boot-strapping method. We conclude that the sample path generation method of CGMYSV model performs well, and it can be successfully applied to American option pricing with LSM. Finally, we present Asian and Barrier option pricing examples with MCS method using the sample path generation algorithm.

~\\
\textbf{Acknowledgments} The author is grateful to Professor Svetlozar T. Rachev for his valuable discussion on this problem and encouragement. The author gratefully acknowledges the support of GlimmAnalytics LLC and Juro Instruments Co., Ltd.

\singlespacing
\bibliographystyle{decsci_mod}
\bibliography{refs_aaron_CGMYSV}
\clearpage

\begin{table}[t]
\centering
\begin{tabular}{ccc}
\hline
 & KS statistic & $p$-value \\
\hline
10 days & 0.0070 & 0.7029 \\ 
25 days & 0.0122 & 0.1026 \\ 
50 days & 0.0069 & 0.7296 \\ 
100 days & 0.0110 & 0.1748 \\ 
\hline
\end{tabular}
\caption{\label{Table KS Test}KS test for distributions of simulated sample paths }
\end{table}

\begin{table}[t]
\centering
\begin{tabular}{ccc}
\hline
Parameter & Call & Put \\
\hline
$\alpha$ & $0.5184$ & $0.0089$ \\ 
$\lambda_+$ & $ 25.4592$ & $2.0852$ \\
$\lambda_-$ & $4.6040$ & $6.2380$ \\
$\kappa$ & $1.0029$ & $1.4333$\\
$\eta$ & $0.0711$ & $0.1961$  \\
$\zeta$ & $0.3443$ & $1.1931$\\
$\rho$ & $-2.0283$ &$-0.1695$\\
$v_0$ & $0.006381$ & $0.0619$\\
\hline
\end{tabular}
\caption{\label{Table:CalibrationEuropeanCallPut}Calibrated Parameters for the SPX option at September 11, 2017.}
\end{table}

\begin{table}[t]
\centering
\begin{tabular}{cccccc}
 \hline
  & \multicolumn{2}{c}{Call} & & \multicolumn{2}{c}{Put} \\ 
  \cline{2-3} \cline{5-6}
 Error & FFT & MCS & & FFT & MCS \\
 \hline
 AAE & $0.4442$ & $0.6220$ & & $0.4047$ & $0.5512$ \\ 
 APE & $0.0053$ & $0.0074$ & & $0.0226$ & $0.0308$ \\
 ARPE & $0.0711$ & $0.0813$ & & $0.2094$ & $0.1672$ \\
 RMSE & $0.6016$ & $0.8019$ & & $0.7006$ & $0.8608$ \\
 \hline
 \end{tabular}
\caption{\label{Table:Errors MCS vs FFT}Error estimators for the parameter calibration to the call and put option market price at September 11, 2017.}
\end{table}

\begin{table}[t]
\centering
\begin{tabular}{cccccc}
\hline
 & \multicolumn{2}{c}{Call} & & \multicolumn{2}{c}{Put} \\
  \cline{2-3} \cline{5-6} 
Number of Simulation & Price & Standard error & & Price & Standard error \\
\hline
 100 & 15.9651 & 2.1001  & &  34.3371 & 7.9356 \\ 
 1000 & 17.8790 & 0.7511 & & 32.5601 & 2.2353 \\ 
 5000 & 19.6536 & 0.3634 & & 32.7791 & 1.0458 \\ 
 10000 & 19.6840 & 0.2551 & & 32.6914 & 0.7617  \\ 
\hline
FFT Price & 19.6590  & & & 32.9541 & \\
Market Price & 19.05  & & & 31.50 & \\
\hline
\end{tabular}%
\caption{\label{Table:ATM}MCS prices and standard errors for SPX call and put with the strike price $K = 2500$ and time to maturity $T= 28\, days$ using calibrated parameters at September 11, 2017.}
\end{table}

\begin{sidewaystable}
\centering
\begin{tabular}{ccccccccccccccc}
\hline
 & \multicolumn{4}{c}{CGMY} & & \multicolumn{8}{c}{CGMYSV} \\
  \cline{2-5} \cline{7-14}
Date & $\alpha$ & $C$ & $\lambda_+$ & $\lambda_-$ & & 
$\alpha$ & $\lambda_+$ & $\lambda_-$ & 
$\kappa$ & $\eta$ & $\zeta$ & $v_0$ & $\rho$  \\
\hline
 Apr. 6, 2016 & 
 $0.5459$ & $0.3495$ & $6.3595$  & $7.7563$ & & 
 $1.1356$ & $35.2115$ & $7.7883$ & 
 $1.9322$ & $0.3550$ & $1.2211$ & $0.0100$ & $1.2237$  \\ 
 Mar. 9, 2016  & 
 $1.6885$ & $0.0246$ & $21.2919$  & $2.8805$ & & 
 $0.0353$ & $20.3911$ & $9.9256$ & 
 $2.1262$ & $0.4570$ & $1.2665$ & $0.0236$ & $1.3434$  \\ 
 Feb. 10, 2016  & 
 $0.9287$ & $3.5284$ & $72.5855$  & $46.4594$ & & 
 $0.0257$ & $1.3491$ & $0.7503$ & 
 $2.4301$ & $0.4185$ & $1.8126$ & $0.2711$ & $0.3567$  \\ 
 Jan. 6, 20166 & 
 $1.2290$ & $0.4104$ & $64.6344$  & $21.3096$ & & 
 $1.3081$ & $33.2233$ & $32.5605$ & 
 $2.0656$ & $0.4237$ & $1.8947$ & $0.0810$ & $0.1172$  \\ 
 Dec. 9, 2015 & 
 $1.3241$ & $0.2984$ & $57.6574$  & $30.2749$ & & 
 $0.6308$ & $40.5190$ & $24.3874$ & 
 $5.0247$ & $0.3299$ & $3.5989$ & $0.0593$ & $0.2334$  \\ 
 Nov. 10, 2015 & 
 $1.5907$ & $0.0184$ & $26.4398$  & $2.7714$ & & 
 $0.0266$ & $1.6365$ & $0.7506$ & 
 $1.0354$ & $0.5658$ & $0.3936$ & $0.1111$ & $1.2914$  \\ 
 Oct. 7, 2015  & 
 $0.9792$ & $0.2955$ & $53.8624$  & $8.4758$ & & 
 $0.8763$ & $65.0630$ & $67.4571$ & 
 $1.8555$ & $0.2794$ & $2.1050$ & $0.0422$ & $0.0080$  \\ 
 Sep. 9, 2015 & 
 $1.2572$ & $0.5862$ & $54.0807$  & $15.3905$ & & 
 $0.5836$ & $30.2731$ & $17.2115$ & 
 $5.1193$ & $0.3569$ & $4.8231$ & $0.1189$ & $0.1158$  \\ 
 Aug. 12, 2015 & 
 $0.7574$ & $0.5926$ & $61.7876$  & $14.2761$ & & 
 $0.4848$ & $42.9235$ & $31.7027$ & 
 $2.1644$ & $0.1964$ & $2.1032$ & $0.0283$ & $0.1639$  \\ 
 Jul. 8, 2015 & 
 $1.1742$ & $0.2429$ & $78.6224$  & $11.4198$ & & 
 $0.9127$ & $46.3671$ & $46.3013$ & 
 $2.2125$ & $0.2173$ & $2.0394$ & $0.0547$ & $-0.0637$  \\ 
 Jun. 10, 2015  & 
 $1.3849$ & $0.0407$ & $64.9276$  & $7.8355$ & & 
 $0.0391$ & $1.0381$ & $0.7820$ & 
 $1.1451$ & $0.6167$ & $0.6062$ & $0.0518$ & $1.2736$  \\ 
 May. 6, 2015 & 
 $1.2247$ & $0.1125$ & $87.0744$  & $8.6069$ & & 
 $0.8628$ & $53.8182$ & $54.3726$ & 
 $2.0949$ & $0.2023$ & $2.0731$ & $0.0352$ & $0.0647$  \\ 
 \hline
\end{tabular}%
\caption{\label{Table:CalibrationAmericanOption}Parameter Calibration Results for the OEX Option market}
\end{sidewaystable}
  
\begin{table}
\centering
\begin{tabular}{cccccc}
\hline
Date & Model & 
AAE & APE & ARPE & RMSE \\
 \hline
\hline
 Apr. 6, 2016 
& CGMY 
& 0.6623& 0.1726& 0.4587& 0.8833\\ 
& CGMYSV 
& \textbf{0.1881}& \textbf{0.0490}& \textbf{0.1242}& \textbf{0.2497}\\ 
 \hline 
 Mar. 9, 2016 
& CGMY 
& \textbf{0.6780}& \textbf{0.1106}& 0.2871& 0.8323\\ 
& CGMYSV 
& 0.6833& 0.1115& \textbf{0.2492}& \textbf{0.8086}\\ 
 \hline 
 Feb. 10, 2016 
& CGMY 
& \textbf{0.7755}& \textbf{0.0513}& \textbf{0.1282}& \textbf{1.1076}\\ 
& CGMYSV 
& 0.9267& 0.0613& 0.2075& 1.2042\\ 
 \hline 
 Jan. 6, 2016 
& CGMY 
& 0.6945& 0.0644& 0.1617& 0.9379\\ 
& CGMYSV 
& \textbf{0.4140}& \textbf{0.0384}& \textbf{0.0784}& \textbf{0.6574}\\ 
 \hline 
 Dec. 9, 2015 
& CGMY 
& 0.7986& 0.0931& 0.1855& 1.1230\\ 
& CGMYSV 
& \textbf{0.5903}& \textbf{0.0688}& \textbf{0.0957}& \textbf{0.7885}\\ 
 \hline 
 Nov. 10, 2015 
& CGMY 
& 0.3137& 0.0576& \textbf{0.2169}& 0.4159\\ 
& CGMYSV 
& \textbf{0.2337}& \textbf{0.0429}& 0.2560& \textbf{0.3815}\\ 
 \hline 
 Oct. 7, 2015 
& CGMY 
& 0.5268& 0.1018& 0.3741& 0.6891\\ 
& CGMYSV 
& \textbf{0.2186}& \textbf{0.0423}& \textbf{0.1520}& \textbf{0.3221}\\ 
 \hline 
 Sep. 9, 2015 
& CGMY 
& 0.8948& 0.0880& 0.1681& 1.2499\\ 
& CGMYSV 
& \textbf{0.5874}& \textbf{0.0577}& \textbf{0.1002}& \textbf{0.9153}\\ 
 \hline 
 Aug. 12, 2015 
& CGMY 
& 0.6464& 0.0941& 0.1894& 0.9097\\ 
& CGMYSV 
& \textbf{0.4137}& \textbf{0.0602}& \textbf{0.1401}& \textbf{0.5652}\\ 
 \hline 
 Jul. 8, 2015 
& CGMY 
& 0.7636& 0.0800& 0.1456& 1.0079\\ 
& CGMYSV 
& \textbf{0.2518}& \textbf{0.0264}& \textbf{0.1023}& \textbf{0.3238}\\ 
 \hline 
 Jun. 10, 2015 
& CGMY 
& \textbf{0.2535}& \textbf{0.0671}& \textbf{0.2576}& \textbf{0.3483}\\ 
& CGMYSV 
& 0.3180& 0.0841& 0.3676& 0.4120\\ 
 \hline 
 May. 6, 2015 
& CGMY 
& 0.7361& 0.0765& 0.1479& 0.9776\\ 
& CGMYSV 
& \textbf{0.3880}& \textbf{0.0403}& \textbf{0.1097}& \textbf{0.5137}\\ 
 \hline 
 \hline
\end{tabular}
\caption{\label{Table:4errors}Error Estimates for the calibration of the OEX option}
\end{table}

\begin{table}
\centering
\begin{tabular}{ccc}
\hline
 &  \multicolumn{2}{c}{Put} \\
  \cline{2-3} 
Number of Simulation & Price & Standard error \\
\hline
 100 &  15.4565 & 2.7280 \\ 
 1000 &  15.0373 & 1.0173 \\ 
 5000 &  13.9616 & 0.3990 \\ 
 10000 &  13.8768 & 0.2856 \\  
   \hline
   Market Price & 13.950 \\
   \hline
\end{tabular}%
\caption{\label{Table:OexMCSExample}MCS prices and standard errors for the OEX put with time to maturity of $T = 31 days$ and strike price $K = 910$ using parameters calibrated at April 6, 2016. }
\end{table}

\begin{table}
\centering
\begin{tabular}{cccccc}
\hline
 & \multicolumn{2}{c}{Call} & & \multicolumn{2}{c}{Put} \\
  \cline{2-3} \cline{5-6} 
Number of Simulation & Price & Standard error & & Price & Standard error \\
\hline
 100 & 22.0834 & 1.7485 & & 11.7400 & 3.4463 \\ 
 1000 & 21.0078 & 0.5866 & & 9.7009 & 1.0759 \\ 
 5000 & 21.4664 & 0.2785 & & 10.5634 & 0.5443 \\ 
 10000 & 21.6513 & 0.1937 & & 9.9964 & 0.3679 \\ 
\hline
\end{tabular}%
\caption{\label{Table:Asian}MCS prices and standard errors for Asian call \& put.}
\end{table}

\begin{table}
\centering
\begin{tabular}{cccccc}
\hline
 & \multicolumn{2}{c}{Down \& Out Call} & & \multicolumn{2}{c}{Up \& Out Put} \\
  \cline{2-3} \cline{5-6} 
Number of Simulation & Price & Standard error & & Price & Standard error \\
\hline
  100 & 12.9019 & 1.4535 & & 36.2543 & 3.6254 \\ 
 1000 & 15.3738 & 0.5165 & & 30.0687 & 0.9509 \\ 
 5000 & 16.6097 & 0.2460 & & 31.8030 & 0.4498 \\ 
 10000 & 16.5518 & 0.1749 & & 30.2590 & 0.3026 \\ 
\hline
\end{tabular}%
\caption{\label{Table:Barrier}MCS prices and standard errors for the down-and-out call and the up-and-out put.}
\end{table}

\clearpage

\clearpage

\end{document}